%% file: outrigger_overview.tex
\documentclass[preprint]{aastex63}

\usepackage{amssymb}
\usepackage{amsmath}
\usepackage[T1]{tipa}
\usepackage{color}
\usepackage{graphicx}
\usepackage{hyperref}
\usepackage{natbib}
\usepackage[capitalise]{cleveref}
\usepackage{multirow}
\usepackage{threeparttablex}
\usepackage{booktabs}
\usepackage{marginnote}
\usepackage{blkarray}
\usepackage{verbatim}
\usepackage{makecell}
\usepackage{wrapfig}
\usepackage[normalem]{ulem}

\DeclareTextFontCommand{\textipa}{%
  \fontfamily{cmss}\tipaencoding
}

\newcommand{\kkoname}{k'ni\textipa{P}atn k'l$\left._\mathrm{\smile}\right.$stk'masqt}

\newcommand{\SNR}{\ensuremath{\text{S/N}}}

\definecolor{rosewood}{rgb}{0.4, 0.0, 0.04}

\begin{document}

\title{CHIME/FRB Outriggers: Design Overview}

\shorttitle{CHIME/FRB Outriggers Overview}
\shortauthors{The CHIME/FRB Collaboration}

\collaboration{1000}{The CHIME/FRB Collaboration}
\input{authors.tex}

\correspondingauthor{Kiyoshi W. Masui}
\email{kmasui@mit.edu}

\begin{abstract}
The Canadian Hydrogen Intensity Mapping Experiment (CHIME) has emerged as the
world's premier facility for studying fast radio bursts (FRBs) through its
fast transient search backend CHIME/FRB\@. The CHIME/FRB Outriggers project
will augment this high detection rate of 2--3 FRBs per day with the ability
to precisely localize them using very long baseline interferometry (VLBI).
Using three strategically located stations in North America and deploying
recently developed synoptic VLBI observing techniques, the Outriggers will
provide $\sim 50$~milliarcsecond localization precision for the majority of
detected FRBs. This paper presents an overview of the design and
implementation of the Outriggers, covering their geographic distribution,
structural design, and observational capabilities. We detail the scientific
objectives driving the project, including the characterization of FRB
populations, host galaxy demographics, and the use of FRBs as cosmological
probes. We also discuss the calibration strategies available to mitigate
ionospheric and instrumental effects, ensuring high-precision localization.
With two stations currently in science operations, and the third in
commissioning, the CHIME/FRB Outriggers project is poised to become a
cornerstone of the FRB field, offering unprecedented insights into this
enigmatic cosmic phenomenon.
\vspace{10mm} 
\end{abstract}


\section{Introduction}
\label{s:intro}
Fast radio bursts (FRBs) are brief and intense flashes
of radio waves originating
from galaxies beyond the Milky Way, with durations ranging from microseconds to
milliseconds \citep[for comprehensive reviews
see][]{2019A&ARv..27....4P,2022A&ARv..30....2P}. Their high luminosities and cosmological
distances make them compelling subjects for astrophysical research. Despite
their intriguing nature, the origins of FRBs remain largely enigmatic, although
they are believed to involve highly energetic processes. In addition to being
interesting in their own right, FRBs serve as probes of the intergalactic and circumgalactic
media and have the potential to unveil the properties of their host galaxies,
the distribution of matter in the universe, and the fundamental physics of the
extreme environments in which they are generated.

The Canadian Hydrogen Intensity Mapping Experiment \citep[CHIME;][]{2022ApJS..261...29C} 
and its fast
transient search backend CHIME/FRB \citep{2018ApJ...863...48C}
currently have the highest detection rate among instruments searching for FRBs,
2--3 per day. CHIME/FRB's first catalog
of 536 events detected in its first year of operations contains the majority of
FRBs reported in the literature to date \citep{2021ApJS..257...59C}. This high
detection rate has produced a number of observational results that have
substantially advanced our understanding of the FRB
phenomenon.

Precise localizations of FRBs are crucial for understanding their origins and
the environments from which they emanate. Localizing an FRB is necessary for 
making an association between a source and its host galaxy, thereby enabling
studies of the demographics and stellar populations of FRB hosts.
Obtaining distances to FRBs---usually via optical redshifts of their hosts---facilitates studies of their
energetics and volumetric abundances \citep{2022MNRAS.509.4775J,2022MNRAS.510L..18J}.
CHIME is capable of localizing FRBs to the arcminute-scale precision
\citep{2021ApJ...910..147M}, which is sufficient to identify host galaxies only
in rare cases of nearby FRBs in $\sim L^*$ galaxies
\citep[e.g.,][]{2021ApJ...919L..24B,2023arXiv231010018B,2023ApJ...950..134M, 2024ApJ...961...99I}.
Consistently identifying host galaxies beyond $\sim 100$ Mpc requires arcsecond localization
precision, which to date has been achieved for several dozen FRBs
by interferometers such as the Karl G. Jansky Very Large Array
\citep[VLA; e.g.,][]{2017Natur.541...58C, 2020ApJ...899..161L}; the Australian Square Kilometre Array
Pathfinder \citep[ASKAP;
e.g.,][]{2019Sci...365..565B,2020Natur.581..391M,2022AJ....163...69B};
the Deep Synoptic Array \citep[DSA; e.g.,][]{2019Natur.572..352R,2024ApJ...967...29L}; and MeerKAT
\citep[e.g.,][]{2024MNRAS.527.3659D}.

For a small number of repeating FRBs, milliarcsecond-scale localizations have been
obtained using very long baseline interferometry (VLBI) with the European VLBI
Network (EVN). Although these instances are few, the
precision of these localizations has provided rich information about the
environments of FRB sources \emph{within their hosts}.
This includes the localization of FRB~20121102
within a dwarf galaxy and coincident with a persistent radio source
\citep{2017ApJ...834L...8M,2017ApJ...834L...7T}; FRB~20180616,
which lies 60\,pc offset from a star-forming region within a spiral
galaxy \citep{2020Natur.577..190M,2021ApJ...908L..12T}; and FRB~20200120E, which resides in a
globular cluster in the halo of M81 \citep{2021ApJ...910L..18B,2022Natur.602..585K}.

Unfortunately, VLBI localizations of non-repeating FRBs face significant
challenges with existing observatories. The field of view of observatories
within VLBI networks is not large enough to detect significant numbers
of FRBs in an untargeted survey, making VLBI localizations only possible in targeted follow-up, which
is impractical for non-repeaters. Morphological differences in the dynamic
spectra of FRBs suggest that
repeating and non-repeating sources may form distinct populations
\citep{2021ApJ...923....1P, 2023ApJ...947...83C, CurtinMorphology2024}.
Consequently,
observations that shed light on the $\lesssim 3\%$ of sources observed to
repeat~\citep{2023ApJ...947...83C}
may not lead to an understanding that applies to the non-repeating population.

FRBs have the potential to serve as a powerful cosmological probe
of the large- and intermediate-scale plasma
distribution \citep{mcq14, ms15, 2019PhRvD.100j3532M, 2020Natur.581..391M,
2025arXiv250117922M}.
Roughly $90\%$ of the Universe's baryonic matter is
in the form of diffuse plasma between and surrounding galaxies.
FRBs can be used to trace this plasma via dispersion, which provides a precise
measure of the free election column density.
In contrast to other probes, which are sensitive only
to the hottest or densest regions, dispersion traces even the most diffuse
plasma.
However, redshift information is critical to
disentangling distance and density in dispersion measurements. Furthermore,
since such studies are fundamentally statistical measurements of the plasma
distribution, large numbers of localized FRBs are required.

\begin{figure}
    \centering
    \setlength{\lineskip}{0pt} 
    \includegraphics[width=0.495\linewidth]{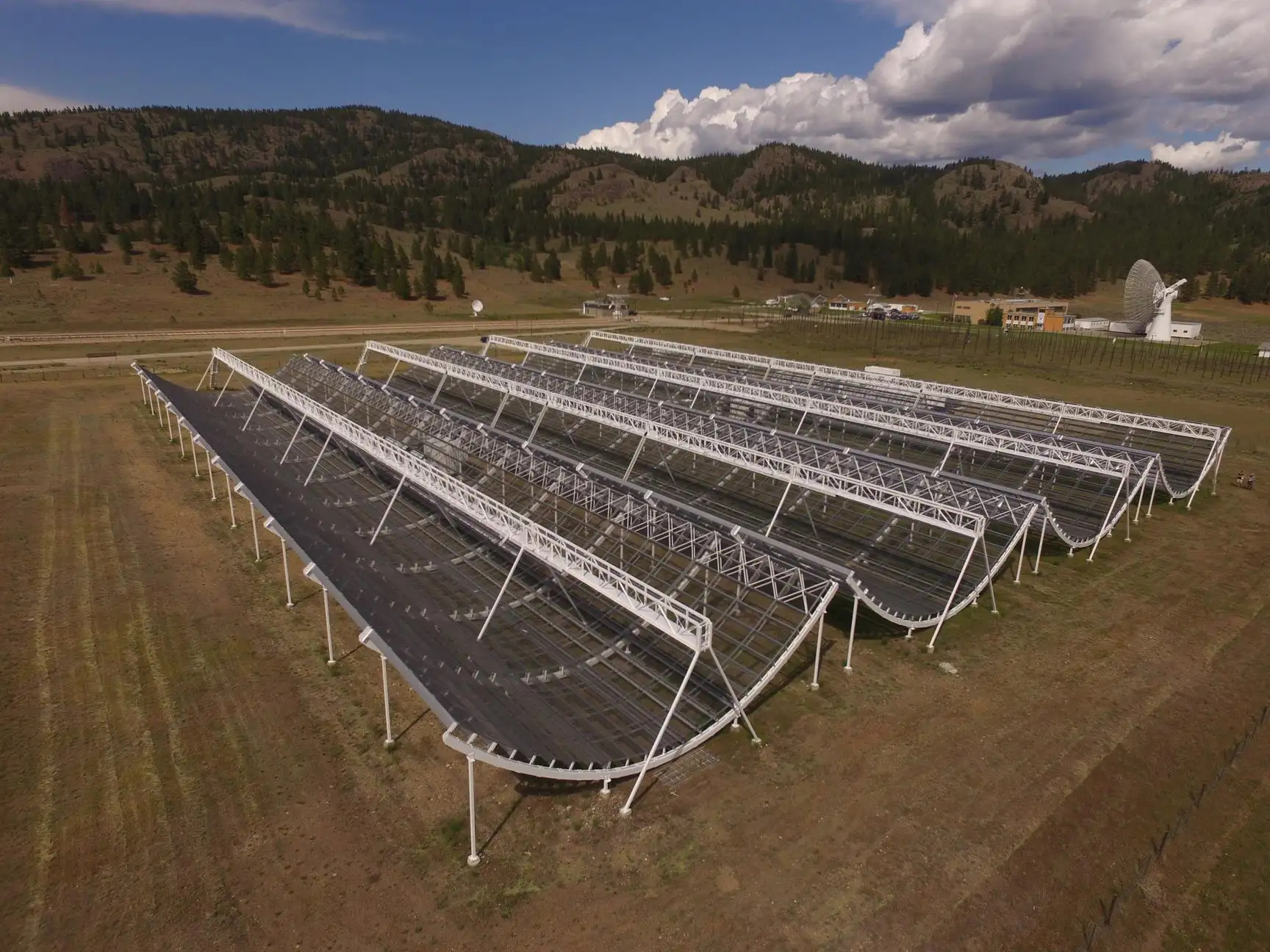}\hspace{-1mm}%
    \includegraphics[width=0.495\linewidth]{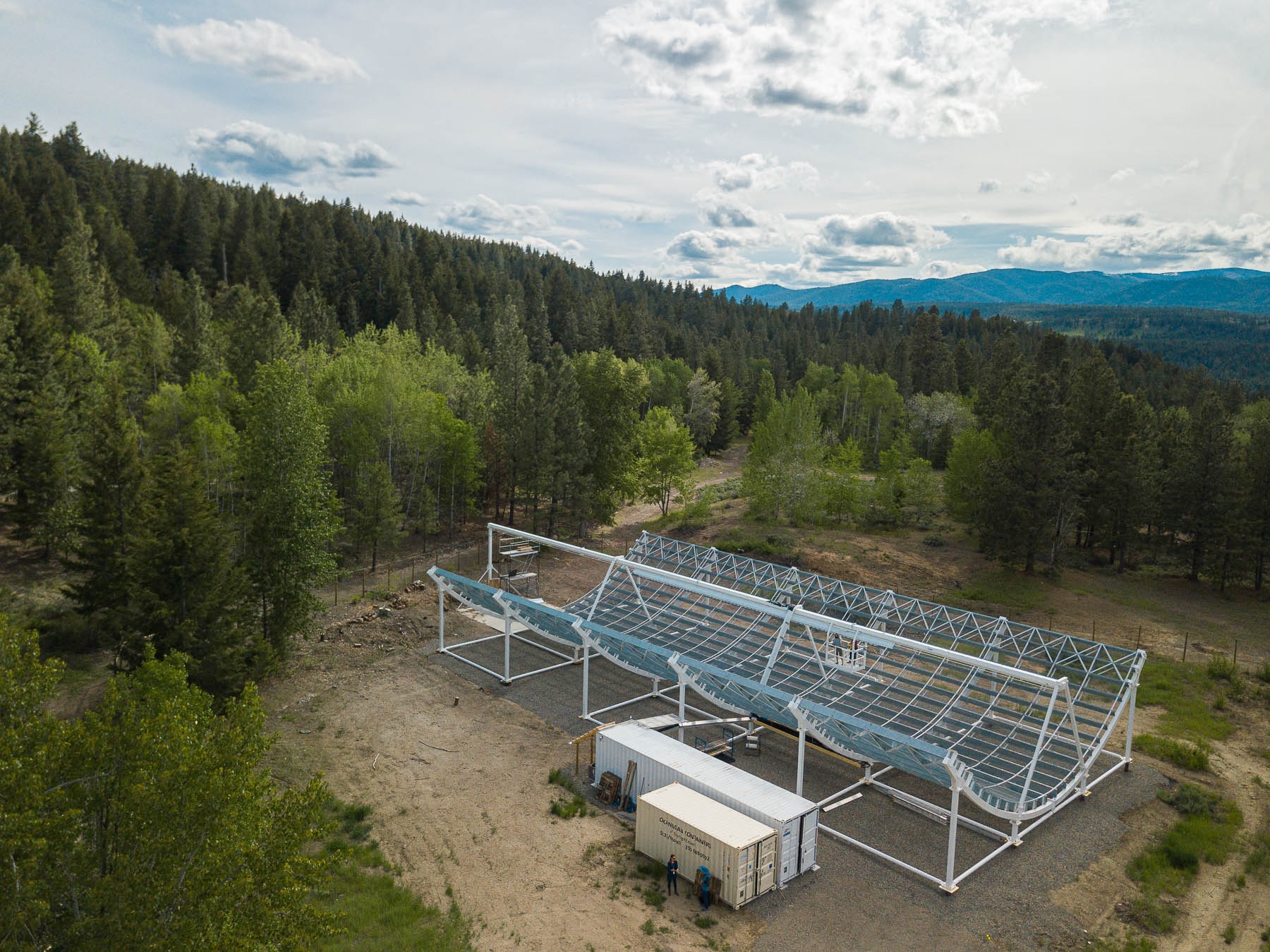}\\[-1mm]
    \includegraphics[width=0.495\linewidth]{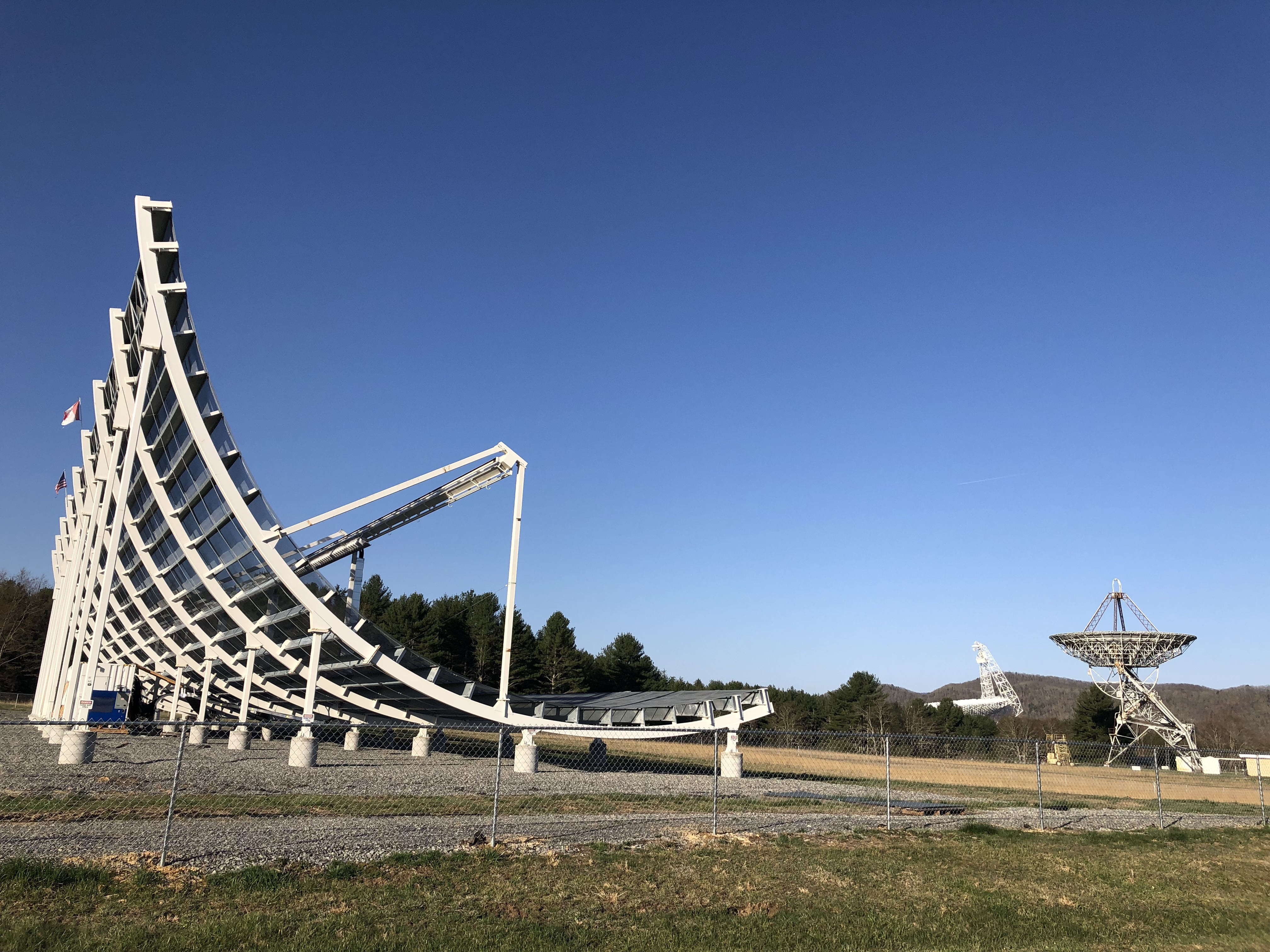}\hspace{-1mm}%
    \includegraphics[width=0.495\linewidth]{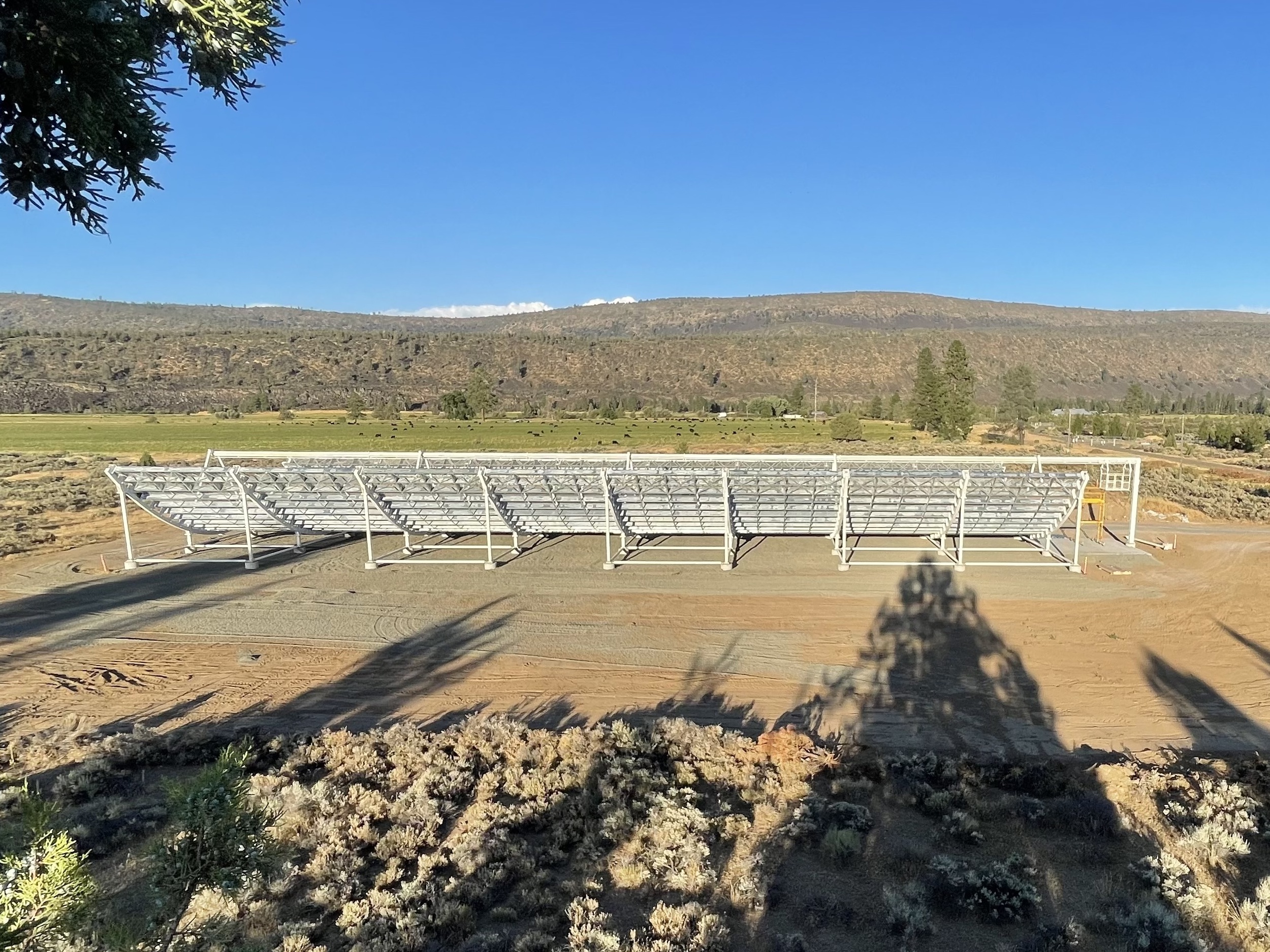}%
    \caption{CHIME and the Outrigger stations. Clockwise from top left: CHIME,
    KKO, and the Outriggers at HCRO and GBO\@. Photo credits: J.~Richard
    Shaw, National Research Council Canada / Conseil national de recherches
    Canada, Kiyoshi Masui, Kenzie Nimmo.
    \label{f:photos}
    }
\end{figure}

Here, we provide an overview of the CHIME/FRB Outriggers project,
which will perform VLBI localizations of thousands of FRBs,
providing the premier dataset for both studying the FRB
phenomenon and deploying them as cosmological probes.
The Outriggers, shown in Figure~\ref{f:photos}, will employ recently developed synoptic VLBI techniques
\citep{2021AJ....161...81L,2022AJ....163...65C,2023arXiv230410534S,2023arXiv230709502C}
to obtain precise localizations for the
majority of FRBs detected by CHIME over its approximately 200~sq.~deg.\ 
field of view.
\citet{2024AJ....168...87L} provided an in-depth overview of the
\kkoname{}
Outrigger\footnote{The name of the first Outrigger \kkoname{} was a generous gift from the
Upper Similkameen Indian Band and means ``a listening device for outer space.''}
(KKO),
the first of the three stations, focusing on the
detailed design and instrument commissioning and performance. Here we take a
broader view and focus on the design and capabilities of the full outrigger
network. The commissioning and performance of the stations at the Green Bank
Observatory (GBO) and the Hat Creek Radio Observatory (HCRO) will be described
in other work.

The paper is organized as follows: Section~\ref{s:objectives} describes our
scientific objectives and translates these into technical
requirements. Section~\ref{s:design} provides a description of the design
of both the full VLBI array as well as the individual telescope stations.
Section~\ref{s:capabilities} provides a description of our observational
capabilities in the context of enabling calibration that will propel us to
achieve our objectives. Section~\ref{s:forecasts} provides statistical forecasts for the
localization performance of the complete array of Outriggers. Finally,
Section~\ref{s:status} describes the current status and outlook for the
program.

\section{Scientific Objectives and Requirements}
\label{s:objectives}

By establishing a large number of ultra-precise localizations, the CHIME/FRB
Outrigger project aims to advance our understanding of FRBS themselves, and to
provide a statistically significant number of probes of Galactic and
extragalactic environments. Our scientific objectives are based on four
pillars.

\begin{description}
    \item[FRB populations] By identifying host galaxies we aim to
        assemble a large sample of FRBs with distances. This will allow for
        precise measurements of their energetics and
        abundances, shedding light on the nature and distribution of FRBs
        across the universe.
    \item[Host galaxy demographics] Through follow-up observations, we intend to characterize the host
        galaxies of FRBs in terms of their global properties, including mass
        and star formation history. This will provide insights into the types
        of galaxies that harbor FRBs and the potential relationship between
        FRBs and their host galaxies' evolutionary stages.
    \item[Source environments] Our goal is to study the specific locations
        within their host galaxies where FRBs occur. To this end, we desire a
        large sample for statistical studies of properties such as offsets
        from galaxy centroids with kiloparsec precision, as well as a smaller
        sample at the lowest possible redshifts for
        detailed, high-resolution multi-wavelength follow-up observations
        for studies of the host environment on scales of $\sim 10$\,pc.
    \item[FRB as cosmological probes] By obtaining a large sample of FRBs with
        accurately determined redshifts over a wide range of redshifts, we plan to
        use FRBs as probes for
        cosmological studies. This includes investigating the large-scale
        distribution of baryons in the Universe, which could offer new insights
        into the structure and evolution of the cosmos.
\end{description}


To meet the scientific objectives we have identified
several key survey and/or technical requirements:

\begin{enumerate}
    \item \textbf{A sample of more than 1000 precisely localized FRBs:} The scope of our scientific
        objectives necessitates a large sample size. For statistical studies,
        the precision of measurements improves with the size of the sample. A
        larger dataset also increases the likelihood of encountering rare and
        illuminating systems that can provide deeper insights into the FRB
        phenomenon. Cosmological studies, in particular, stand to benefit
        significantly from a large sample size, enabling precision measurements
        with substantial scientific impact. With the current tally of
        high-probability FRB hosts exceeding 100 at the time of writing,
        a tenfold increase to 1000 hosts will
        represent a major advance in our statistical power.
    \item \textbf{Localization precision of $\lesssim50$\,mas:} 
        For typical CHIME-detected FRBs at 
        gigaparsec distances, 50 mas corresponds to sub-kiloarsec physical resolution,
        sufficient for statistical studies of source locations.
        For nearby events at $\lesssim 100$\,Mpc distances, the physical
        scale of the localization precision
        becomes $\sim10$\,pc, providing information on the location within, e.g.,\
        star forming regions or offsets from globular clusters and supernova
        remnants. This target is also well
        matched to the highest resolution instruments at other wavelengths,
        e.g.,\ the Hubble Space Telescope ($\sim100$\,mas), Keck adaptive optics ($\sim 10$\,mas)
        JWST ($\sim100$\,mas), 
        the Atacama
        Large Millimeter Array ($\sim20$\,mas), and upcoming telescopes like
        the Vera C. Rubin Observatory ($200$\,mas) and extremely large
        telescopes ($\sim10$\,mas). 
        Finally, our target localization precision strikes a balance
        between technical feasibility (which is discussed in more detail
        throughout this paper) and the potential for scientific
        discovery, particularly for FRBs in the nearby Universe.
    \item \textbf{High completeness or well-understood selection functions for
        host identifications and redshifts:} Accurate characterization of FRB
        populations and their host galaxies is paramount. Selection effects can
        significantly bias these characterizations, and the impact of such
        biases on cosmological studies is not fully understood. The CHIME/FRB
        project characterizes the radio selection function 
        of the FRB search engine using a real-time system of in-situ
        synthetic pulse injections \citep{2023AJ....165..152M}.
        To measure property distributions (of
        either FRBs or their hosts) at 10\% precision, host identification and redshift
        determination must be within a few percent of being complete, or the
        selection function must be either measured or modelled at the few
        percent level.
\end{enumerate}

Our requirement (3) of knowing the selection effects merits further
discussion. After having met our localization precision requirements (2), our
completeness will be limited not by the radio localization but by the ability
to detect the host in the optical band and obtain its redshift. Since
such factors feed more into our follow-up program than
Outrigger design, we defer their
detailed discussion to future work (Andersen et al.~2025, in preparation).
Thus the radio
localization will not be a source of incompleteness, so long as we meet our
astrometric precision requirements for a subset of FRBs for which the selection is
well understood. For example the effects of imposing some \SNR{}
cutoff can be calibrated via synthetic signal injections into CHIME/FRB's 
detection pipeline \citep{2023AJ....165..152M}.

It is also acceptable to include only a subset of the sky in our completeness
requirement, so long as the included sky fulfills our sample-size requirement
(1). At a minimum, it will be necessary to exclude dusty sightlines that preclude optical
spectroscopic follow-up. Existing optical surveys (either imaging or
spectroscopic galaxy surveys) can substantially reduce the required follow-up
within those fields, and can be used to calibrate the completeness in other parts of
the sky.

In practice, we expect our selection effect requirement to be fulfilled through
a combination of the maximum possible completeness in both the radio and
optical, and modelling
the remaining incompleteness to enough precision such that the residual
uncertainty is at the few percent level. Even so, selection issues will arise
if completeness strongly depends on some aspect of the population that we wish to
study. For example, repeating FRBs are known to be preferentially narrowband
compared to non-repeating FRBs \citep{2021ApJ...923....1P, 2023ApJ...947...83C}, 
and as we show in Section~\ref{s:forecasts},
narrowband FRBs are harder to localize. As such, effort will be required to
be complete to narrowband FRBs and throughout the FRB parameter space.

\section{Design}
\label{s:design}
The design of the CHIME/FRB Outriggers was driven by the
scientific objectives specified in the preceding section. Here, we explain the
rationale for our design choices, which divide into two parts:
the design of the VLBI array, through the
geographic locations of the stations, and the design of the stations
themselves.

\subsection{Geographic Array Design}
\subsubsection{Baseline length scale}
The scientific objectives listed in the previous section motivate localization
precision in the 10 to 100\,mas range. On a single baseline, the
statistical limit to the localization precision is \citep{1970RaSc....5.1239R}
\begin{equation}
    \sigma_\theta^\text{stat} = \frac{c}{2 \pi\,b\,(\SNR_\times)\,\text{BW}_\text{eff}},
\end{equation}
where $\sigma_\theta^\text{stat}$ is the localization angle statistical error,
$b$ is the baseline length, $\SNR_\times$ is the signal-to-noise ratio in VLBI
cross-correlation, and $\text{BW}_\text{eff}$ is the effective bandwidth
(approximately equal to recorded FRB bandwidth). The factor of $1/(\SNR_\times)$
is a ``super-resolution'' boost, which results from the possibility of measuring phases
to sub-radian precision at high \SNR{}. This is analogous to
optical telescopes being able to measure the centroid of point sources to
precision $\sim\text{FWHM}/(\SNR)$, where FWHM is the full width at half maximum of
the point spread function.
This boost can be difficult to realize beyond 
$\SNR_\times \sim 10$
as systematic uncertainty in phase, for instance, induced by telescope optics,
becomes dominant.
Other significant systematic
errors include the ionosphere and clock calibration.
These will all be discussed further below.

Plugging in CHIME's 400 to 800\,MHz passband,
we have
\begin{equation}
    \sigma_\theta^\text{stat} =
    24\,\text{mas}\left(\frac{10}{\SNR_\times}\right)\left(\frac{100\,\text{km}}{b}\right).
    \label{e:sig_theta_chime}
\end{equation}
As can be seen, baselines in at least the 10 to 100\,km range, depending on our
ability to super-resolve, are required to meet our science goals.
Limitations of CHIME's site at the Dominion Radio Astrophysical Observatory (DRAO) make
the use of
connected-element interferometry at these baselines impractical, instead requiring VLBI.

Since our science goals already pushed us into the VLBI regime, we were faced
with the further question of what baseline length to choose. Stations placed
between 10 and 100 \,km of CHIME see reduced ionosphere variation, and
are more accessible to CHIME personnel, but are capable of providing only a
subset of our science goals, and require developing ``green-field'' sites with
no pre-existing infrastructure.
Longer baselines in the 1000\,km open up many more options for sites,
including cleaner RFI environments and observatories with existing infrastructure
throughout North America. Such baselines will see more ionospheric
variation, however, should systematic errors be controlled,
they have the potential to greatly exceed our localization requirements and
enable ancillary science.
We thus concluded that baselines in the 100 to 1000\,km scale
best fit our requirements.

\subsubsection{Number of stations and distribution}
\label{s:baselines}

Each baseline in an interferometric array
measures a single Fourier mode of the sky. As such, when phrased as an imaging problem, one anticipates needing a
large number of baselines to reconstruct the sky with any fidelity. However,
with the Outriggers we
are pursuing only a localization, not an image, meaning there
are only two numbers of interest per FRB (the 2D sky location). Since FRBs can
be separated from backgrounds in the time domain, it should be possible to
infer these two parameters from only two measurements. However, the location of
a point source cannot be inferred from two Fourier measurements, since each
Fourier mode is minimally compact in sky coordinates, with $2\pi$ phase
ambiguities, resulting in a comb of equally-possible sky locations. This
ambiguity is broken when performing multi-frequency observations, since a fixed
baseline measures a different Fourier mode at each frequency, whereas the
localization is frequency-independent. Put differently, over a finite frequency
range, the visibilities from a point source can be synthesized into a single
geometric \emph{delay} measurement for each baseline, and the 2D point-source
location can be inferred unambiguously from only two delay measurements on
non-parallel baselines. CHIME's broad 400--800\,MHz observing band is
well-suited to not only inferring geometric delays, but separating them from
dispersive delays due to the ionosphere. This process, called fringe-fitting,
is similar to LOFAR's clock-TEC separation procedure \citep{Weeren_2016}, and is
described in more detail in Section~\ref{s:capabilities}.

With each Outrigger site incurring both a substantial financial cost in site
preparation and development, and a large logistical cost for construction,
deployment, and operation, there is a great advantage to minimizing the total
number of stations. The minimal array that fulfills our science goals is two
stations forming non-colinear (and preferably close to orthogonal)
$\sim$1000\,km baselines. However, there were a number of advantages to 
a third station at $\sim$100\,km baseline to CHIME\@. The first is the ability
to do early science since a station that is accessible from CHIME can be
built faster by available team members based at DRAO\@. The single
baseline yields only 1D localizations, but, often this is sufficient for
host-galaxy identification, and has enabled early science
goals \citep{2025arXiv250211217A, 2025ApJ...979L..21S,2025ApJ...979L..22E}.
Second, the
shorter construction timeline enabled refinement of structural and system
design for the other sites, as well as earlier development of observational
capabilities, calibration techniques, and analysis pipelines.
The ionosphere, with a
characteristic height of around 300\,km, is expected to be broadly similar
between stations separated on that scale, providing a means to separately
characterize the performance of our array with and without the ionosphere.
Another advantage is operational redundancy, enabling a sufficiently-precise
localization for host identification and redshift determination if either
of the distant stations is offline.
Finally, a third baseline provides an internal consistency check on FRB
localizations, for which external validations are largely
impossible.

With these considerations, our design comprises one station at a
$\sim100$\,km distance scale from CHIME, and two more at $\sim$1000\,km from
CHIME.

\subsubsection{Site selection}

\begin{figure}
    \centering
    \includegraphics[scale=0.35]{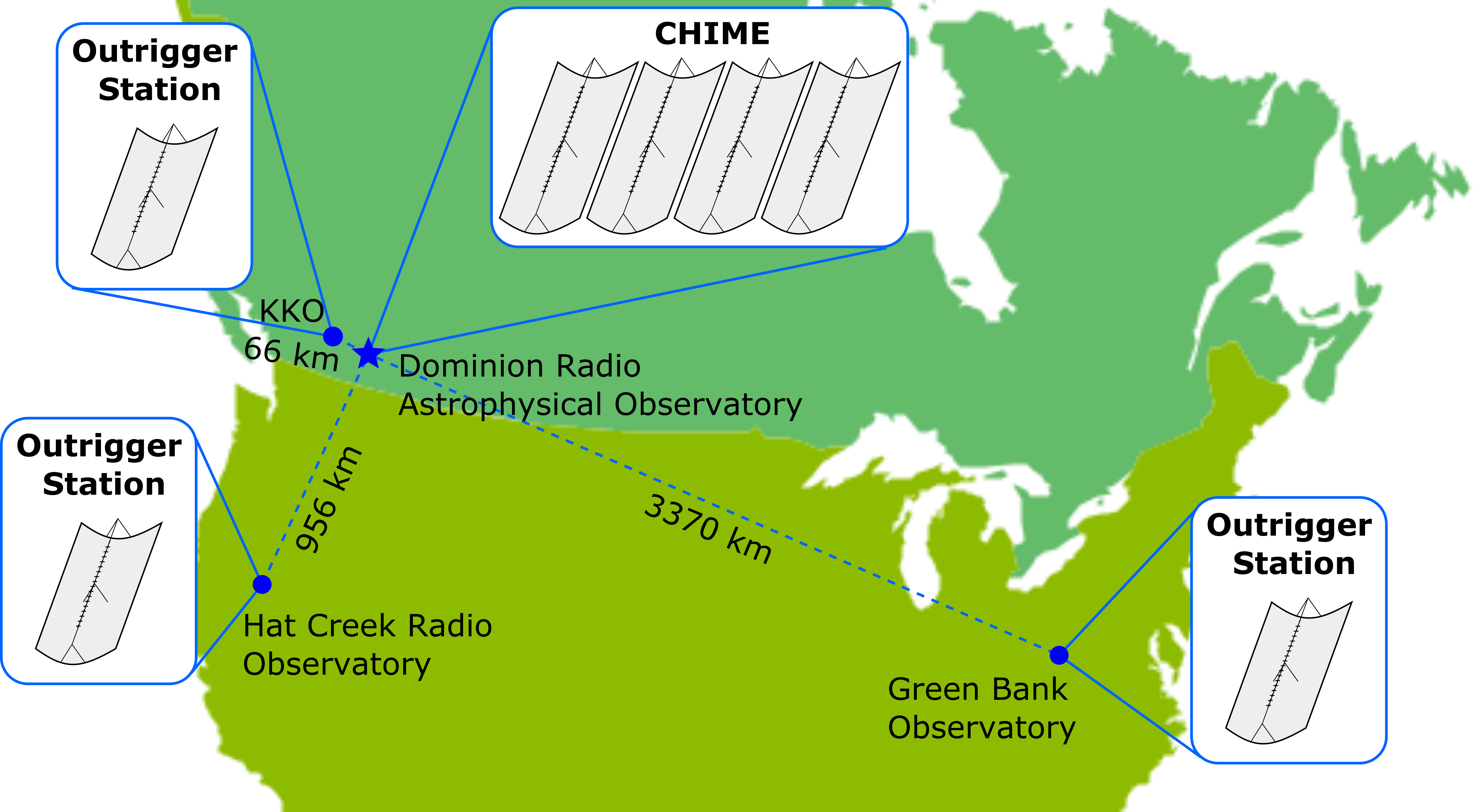}
    \caption{Map of geographical locations of the CHIME and the Outrigger
    sites. Baselines between each Outrigger and CHIME are shown with dotted
    lines with corresponding label of baseline lengths.}
    \label{f:map}
\end{figure}

The map of our selected sites, constituting the full Outriggers VLBI network,
is shown in Figure~\ref{f:map}.

The first site, which hosts KKO, is situated about $\sim 4$~km south of the town of Princeton in
British Columbia. At $66$~km line-of-sight distance to CHIME, it is close
enough to provide easy access to our staff located at DRAO or the University of British Columbia (UBC)\@. This
relatively short baseline allows for arcsec-scale resolution in the baseline direction 
while avoiding the
worst effects of the ionosphere.

The second site selected was at the Green Bank Observatory (GBO) in West
Virginia.
At 3370\,km from CHIME, it forms the longest baseline and so enables the
highest possible precision among the Outriggers. The observatory is in the
National Radio Quiet Zone, reducing radio frequency interference (RFI). Being a
national observatory, the site already has significant infrastructure
including a hydrogen maser clock, which was routed to the Outrigger
backend for precision timing.

For the third site, we selected the Hat Creek Radio Observatory (HCRO) in
Northern California, which provides substantial infrastructure and a radio-quite site.
Its proximity to an airport
with commercial flights in Redding, CA (approximately 90 minute drive) enables convenient access to the science team.
Being 956\,km south of CHIME, the
station provides a long baseline roughly orthogonal to that with GBO\@.
Infrastructure available at HCRO spans heavy construction equipment to
radio-frequency test instruments.

\subsection{Station Design}

Here we provide an overview of the design of the individual Outrigger stations.
We start with the high-level design, including the required collecting area and
best antenna architecture. We then proceed to more detailed aspects, including
elements of the structure, analog chain, digital hardware, and software
pipelines.

\subsubsection{Required collecting area}
\label{s:collecting_area}

The first consideration in designing the Outriggers was the required collecting area for
the stations, driven by the required CHIME--Outrigger cross-correlation
\SNR{} for localization. In the statistically ideal case, this is
given by
\begin{equation}
    \SNR_\times = \SNR_\text{CHIME}\sqrt{\frac{2
    A_\text{eff,out}}{A_\text{eff,CHIME}}}.
    \label{e:snr_times}
\end{equation}
Here, $\SNR_\text{CHIME}$ is the signal-to-noise ratio of the FRB as measured
using offline analysis pipelines in CHIME data alone, and $A_\text{eff,out}$
and $A_\text{eff,chime}$ are the effective
collecting areas of the Outrigger and CHIME respectively. The factor of
$\sqrt{2}$ in
the above equation originates from differences in the statistics in
cross-correlation vs auto-correlation \citep{2015A&C....12..181M}.\footnote{%
One way to understand the factor of $\sqrt{2}$ is to consider the case where the Outriggers
have identical antennas to CHIME, and we treat the feeds in the
station arrays as individual VLBI elements. If CHIME has $n$ feeds and an
Outrigger station $m$ feeds, then CHIME has $\approx n^2/2$ intra-station
baselines, whereas there are $nm$ baselines between CHIME and the Outrigger.
}

Our scientific requirement (1) is that we localize a large number of sources, a
substantial fraction of the $\sim1000$ per year detected by CHIME/FRB\@.
CHIME/FRB's detection threshold is $\SNR_\text{CHIME} > 9$, however, false positives that
are difficult to veto in real time make capturing the requisite baseband data challenging
below $\SNR_\text{CHIME} = 12$ \citep{2024ApJ...969..145C}. While lowering the baseband
capture threshold is likely possible with some development effort, our
experience is that detailed analysis of lower \SNR\ bursts is challenging. In
any case, the higher threshold reduces the total number of bursts by a fraction
$(9/12)^{3/2} = 0.65$ \citep{2021ApJS..257...59C}, which we deem to be an acceptable loss.

We estimate that $\SNR_\times > 6$ is
required to attempt VLBI localization. While this substantially exceeds what is
required for statistical localization precision on 1000\,km baselines as
given in Equation~\ref{e:sig_theta_chime}, 6 is the minimum $\SNR_\times$ for
which we can confidently detect and identify the correct VLBI fringe, as verified
using pulsars and steady sources (discussed in detail below). Note
that in the fringe search there is a non-negligible ``trials factor''
\citep[see Sec. 9.3 of][]{2017isra.book.....T} since the
search space of potential localizations is large (for each baseline, which provides a 1D
localization, this is roughly the ratio of the
CHIME/FRB baseband localization precision to the VLBI localization precision,
which is $\sim 1'/10\,\text{mas} = 6000$). Note that the $\SNR_\text{CHIME}=12$
criterion used for triggering is the \SNR \, yielded by the real time
search. For most events, offline analysis can achieve substantially higher \SNR \,
through coherent dispersion, forming a tied-array beam to the baseband
localization \citep{2024ApJ...969..145C}, and detailed modeling of the burst morphology
\citep{2023arXiv231105829F}. Offline analysis thus build some conservatism into
our collecting area specifications.

Given a $\SNR_\text{CHIME}$ threshold of 12, and a required $\SNR_\times > 6$,
Equation~\ref{e:snr_times} gives
$A_\text{eff,out} = A_\text{eff,CHIME}/8$ as the appropriate station
size. CHIME has an instrumented collecting area of 6400\,m$^2$ and an aperture efficiency
of $\eta_A \approx 0.7$ \citep{2022ApJS..261...29C}, which motivates
$A_\text{eff,out} \approx 550\,\text{m}^2$.

We have so far only discussed CHIME--Outrigger baselines and have not mentioned
Outrigger--Outrigger baselines. For the latter, $\SNR_\times$ is suppressed by
a factor of $\sqrt{A_\text{eff,out}/A_\text{eff,CHIME}}$ compared to the
former. Furthermore, as explained above, only 2 baselines are required to
achieve a 2D localization. As
such, the stations were sized to only rely on the baselines that include
CHIME\@. We anticipate the inter-Outrigger baselines being useful only for the brightest
FRBs or ancillary science.

On a final note, as discussed in Section~\ref{s:baselines}, the specific scientific
objectives of KKO at a baseline of
66\,km differ from the other two sites. KKO's focus on early-science from
a less-complete sample of host identifications can be achieved with a smaller
collecting area. Furthermore, KKO's 1D localizations will enable host
associations only for lower-DM, lower-redshift sources. These preferentially
have higher $\SNR_\text{CHIME}$ \citep{smb+23,2024ApJ...975...75L}. As such, we chose
$A_\text{eff,KKO} = A_\text{eff,CHIME}/16$.

\subsubsection{Antenna architecture}

The required collecting area, $550\,\text{m}^2$, could be obtained from a
number of antenna architectures, but we had the additional requirement that we cover
the CHIME field of view, a 3-degree wide strip
between the North and South horizons over the local meridian at DRAO\@. Several
architectures were considered.

We used CHIME's architecture of 20-m wide cylinders as a baseline, and which was
ultimately selected.
The required collecting area can be achieved with a single
cylinder with 40-m instrumented length.
The major disadvantage of this design is its large monolithic structure, where
the cost of the material, structure fabrication, and foundations dominate.
Furthermore, access to the focal line, 5\,m above the reflector, to install and
service analog components is challenging. A key advantage is that matching
CHIME's field-of-view is straightforward.

Another possibility was a large array of dipoles, as in the Square Kilometre
Array-Low\footnote{\url{https://www.skao.int}}, which would have
a field-of-view covering the visible sky
and would minimize structural costs. This would have been at the cost of requiring an
extreme number (order $10^4$) of analog elements, since each dipole has an 
effective area $\sim\lambda^2/4$. Even if the cost of these would be
reduced to $\sim\$100$ per element, this would be comparable to structural costs
of other architectures. Through analog summing, the number of
digitization channels could be made to be less extreme. Nonetheless, a dipole
array was rejected due to the high analog cost and large departure from
the single-element analog-to-digital chains with which our team has substantial
experience.

An array of dishes with a diameter in the 5 to 10\,m range was also considered.
This would have solved challenges due to a cylinder's monolithic structure, and at smaller
diameters focal points could be accessed via ladder. However, dishes have a
severe mismatch to CHIME in field-of-view. Covering a substantial fraction of
CHIME's FOV could have been achieved by tiling it with effectively-independent arrays of
dishes, but that would increase both costs and the total footprint of the site,
substantially constraining site selection. As such, a dish-based architecture
was rejected.

An option that was carefully considered was a cylindrical architecture but with
a 10\,m width instead of CHIME's 20\,m. Achieving the same collecting area
would have required doubling the instrumented length as well as the number of
analog chains and digitizers. However, structural and foundation costs would
be substantially lower and the focal line, at a height of 2.5\,m, would be
serviceable by ladder.  More detailed costing estimates including structural,
analog, and digital costs suggested that 20\,m and 10\,m widths would be
comparable.

The key advantage of using the same antenna architecture as CHIME is the prospect
of matching beams.  Differences in the primary beam between
elements---particularly in the beam-induced phase---is a major systematic error in
radio interferometry. Matching the beams between elements, in this case between
CHIME and the Outrigger, eliminates this issue. However, even an Outrigger
with identical optics to CHIME would have some mismatch due to zenith angle
differences arising from the different latitudes of the Outrigger sites
Nonetheless, substantial
investment has been made in calibrating and modelling CHIME's beam, so it is
likely that such measurements could be used to correct for any beam
differences.
Ultimately, it was decided that copying CHIME's architecture as closely as
possible with 20-m cylinders was feasible and was the most promising for
mitigating systematic errors.

\subsubsection{Station Design Summary}
\label{s:station_design}

\begin{figure}
    \begin{center}
        \includegraphics[width=0.6\linewidth]{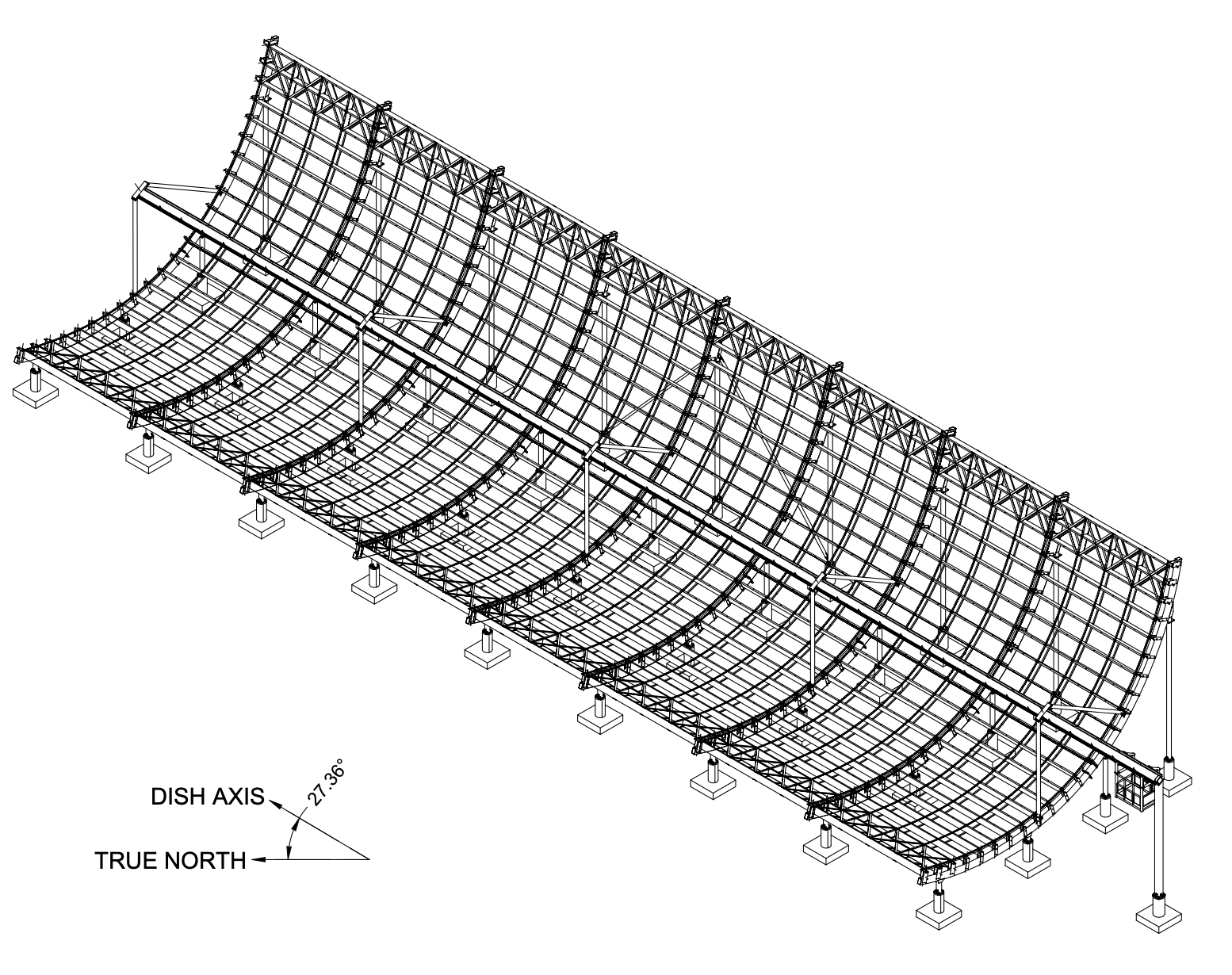}
    \end{center}
    \caption{Isometric view of the structural drawings for the GBO station.
    Due to its Eastward location, the GBO station requires a substantial
    axial roll and a rotation from local North (in the plane of the ground)
    so as to approximately match its field of view to CHIME's along
    the meridian at DRAO\@. Image credit: Sightline Engineering.
    \label{f:gbo_drawing}
    }
\end{figure}

Having settled on 20-m wide cylinders as the station architecture, the
structural designs largely mimic that of CHIME with a few key differences.
Our size requirements from Section~\ref{s:collecting_area} are achieved
for the GBO and HCRO stations with single cylinders with 40\,m of
instrumentation. This becomes a total reflector length of 64\,m when allowing for 10\,m
of buffer reflector at either end (required for proper optics) and when
constructing out of an integer number of 8-m sections, which is the spacing
between the main structural beams.
For KKO, this is 20\,m of instrumentation,
40\,m total. The cylinders are oriented such that they
observe nearly the same field of view as CHIME along the DRAO meridian. This
required the cylinders to be both rolled (along the cylinder axis) and rotated
(in the plane of the local ground). These orientation differences are
particularly pronounced for the station at GBO, due to its site at a
substantially different longitude to CHIME, as can be seen in
Figure~\ref{f:gbo_drawing}. 
For the GBO and HCRO stations, there is a $\sim10\deg$
mismatch in declination due to
the Outriggers being at different latitudes to CHIME\@. This results in
sensitivity mismatches at the few tens of percent level, which is most
pronounced for the most Northerly targets in CHIME's field-of-view.

Furthermore, access to the instrumentation at the
focal lines is from below, in contrast to CHIME where it is from above.
Access is enabled
from a traveling cart that runs on two rails
fixed to the sides of the focal-line structure. Due to design challenges
and safety concerns specific to the high roll of the focal line, the 
cart is no longer in use for the GBO station, where access is now 
provided by aerial lift truck.

Like CHIME, the Outrigger focal lines are instrumented with dual-polarization
clover-antenna feeds \citep{2017arXiv170808521D}, spaced by 305\,mm, for a total
of 64 feeds for KKO and 128 for the other two. Signals from each polarization channel are
amplified by low-noise amplifiers (LNAs) at the feeds. Two such feeds and
associated LNAs are assembled into focal line modules called cassettes, which
also incorporate the ground plane, facilitate deployment, and provide
weatherproofing. In contrast to CHIME, the Outrigger cassettes incorporate a
vacuum-formed acrylonitrile butadiene styrene (ABS) polymer radome to keep
moisture out. Signals are transported by coaxial
cable to the shielded room in the receiver hut, then further amplified and
filtered to the 400 to 800\,MHz observing band by filter amplifiers (FLAs).
There is no frequency conversion as these signals are directly digitized in the
second Nyquist zone. To achieve optimal signal levels for digitization,
additional attenuation or filtering is added depending on the site-dependent
RFI environment.

The Outrigger digital backend is designed as an FX correlator. The F-engine is
implemented using the same field programmable gate array (FPGA)-based ICE
framework used in CHIME \citep{2016JAI.....541005B,2022ApJS..261...29C}. A
single ICE board can digitize and process 16 radio-frequency channels,
or the equivalent of
8 dual-polarization feeds, requiring 16 and 8 boards for the larger Outriggers
and KKO respectively. 
The analog to digital converters (ADCs) on each ICE board are configured to sample the incoming voltage
stream at 800~MS/s. Due to the filtering in the FLAs, this recovers the 400 --
800~MHz operating band of CHIME, corresponding with the second Nyquist zone.
Channelization is done with a polyphase filter bank (PFB) in combination with a
fast Fourier transform (FFT), producing 1024 channels, each 390~kHz wide. The
ICE boards also perform the \textit{corner turn}, regrouping the data by individual frequency channels
instead of by the larger RF channel. These data are henceforth referred to as the
``baseband'' data.\footnote{While the term ``baseband'' is accurate due to the
absence of mixing in CHIME's radio-frequency system, it is not particularly
descriptive of the phase-persevering measurements of the incident electric
field. Nonetheless, this nomenclature is commonly used in some radio-astronomy
sub-fields and has been adopted within the CHIME/FRB Collaboration.}

The frequency-channelized and frequency-grouped baseband data are transmitted
via direct network
links to the \textit{X-engine}.
The X-engine performs three key operations on the data. First is the correlation
the data into the $N^2$ visibilities, which enables each station to
operate as an independent interferometer and facilitates calibration of the
intra-station array. Second, these nodes host a large quantity of random access
memory (RAM) to buffer the full-array baseband data so that it can be
recorded upon receipt of an FRB trigger. This constitutes the primary science
data stream used for VLBI localization, which will be described in more detail
in Section~\ref{s:baseband}. Finally, tied-array beams are formed to targeted
sources, enabling traditional, non-triggered VLBI observations, which will be
used for calibration and ancillary science. This capability is described in
further detail in Section~\ref{s:tracking_beams}.

The X-engine consists of x86-architecture computers
housing graphics processing units (GPUs) for accelerated computing. The number of such nodes is driven by
the input data baseband data rate, with each node designed to receive about 205\,Gbps.
Four such nodes are required for each of the larger Outriggers and two for KKO.

Precise and accurate clocking and time-tagging of the data is a critical
consideration for a VLBI system, where, as will be discussed, localization
can be framed as a precision timing problem. Clock and timing signals for the
F-engine (which hosts the ADCs) is generated by a GPS-disciplined crystal
oscillator, which provides absolute time tagging at tens of nanosecond precision but is
lacking in short and medium-term stability. As described by
\citet{2022AJ....163...48M}, we use
an auxiliary clock, fed directly into a correlator ADC,
to correct our timing on these timescales. At KKO, the auxiliary clock is a 
is a free-running Stanford Research Systems FS725 Rubidium clock. For the GBO station and
CHIME we use the signal from the observatory maser. For the HCRO station we use the
observatory clock which is an Endrun Technologies Meridian II Precision
TimeBase with the US-Rb oscillator upgrade.
This is a GPS-disciplined time and frequency standard whose underlying oscillator is the
same Stanford Research Systems Rubidium clock as KKO\@.
\citet{2022AJ....163...48M} and \citet{2021RNAAS...5..216C}
demonstrated the ability to achieve 0.2\,ns timing precision on 2000\,s
timescales for the stations with rubidium clocks and roughly day timescales for
stations with masers.

\section{Observational Capabilities and Calibration Strategies}
\label{s:capabilities}
In this section we provide an overview of the Outriggers' observational
capabilities and how they will enable us to meet our localization requirements.
We begin with the detailed formalism of the calibration--localization inference
problem, then provide an overview of available calibration sources within our
band, and then proceed to observational capabilities and strategies that enable
us to achieve our science goals.

The ultimate calibration strategy that will be used for the
Outriggers is not finalized. This stems from several sources of underlying
uncertainty including availability of VLBI calibrators in our
band, the spatio-temporal properties of the ionosphere, and what strategies
turn out to be operationally most feasible. Nonetheless, work done so far has
led to a promising calibration strategy where, upon receipt of an
trigger from CHIME/FRB, observations of the nearest suitable pre-cataloged compact continuum calibrators
are initiated at low-latency using coordinated tracking beams at CHIME and each
Outrigger station. This strategy is detailed in
Section~\ref{s:triggered_cal}.

\subsection{VLBI Localization Measurement Equation}
\label{ss:meas_equation}

Here we describe our model for VLBI observations within the context of the
point-source localization problem, including 
how phase referencing can mitigate the effects of unknown phase contributions to the visibility. We then discuss
the parametrics of the measurement, including contributions to the
interferometric phase from the geometric delay
(which contains the localization information), and nuisance parameters for offsets in the
observatory clocks and differences in dispersive delays due to the ionosphere.
Finally, we use this understanding to motivate requirements for calibration
observations that will enable use to achieve our localization goals.

Our model for the signal in a VLBI visibility is as follows:
\begin{align}
    V^s_{ab}(\nu, t)
    &= S^s_\nu \exp\left\{
        i2\pi \nu \left[
            \tau_{ab}( \hat n^s, t)
            + \tau^\text{cl}_{ab}(t)
            + \kappa\,\text{sTEC}_{ab}^s(t) /\nu^2
            \right]
        + i\phi_{ab}(\nu)
        \right\}.
\end{align}
Here, $V^s_{ab}(\nu, t)$ is the visibility signal from source $s$ between
stations $a$ and $b$ as a function of observing frequency $\nu$ and time $t$. We note that
for FRBs, the observation time is near instantaneous, but frequency dependent due to the
dispersed pulsed emission. However, for the time being we keep things general
and carry forward the time and frequency dependence.
$\tau_{ab}( \hat n^s, t)$ is the geometric delay between the two stations for
source sky location $\hat n^s$. $S^s_\nu$ is the source's flux density and 
$\tau^\text{cl}_{ab}(t)$ is the difference in
clock delays between the two stations.
$\text{sTEC}_{ab}^s \equiv \text{sTEC}_{a}^s - \text{sTEC}_{b}^s$ is the
difference in ionospheric slant total electron content between the two
stations
in the direction of the source.  $\kappa$ is a physical constant such that
$\kappa\,\text{sTEC} / \nu^2$ is the dispersive delay from the ionosphere.
We adopt a value of $\kappa = 1.345 \times 10^3~\mathrm{MHz~TECu}^{-1}$
\citep{Weeren_2016}.
$\phi_{ab}(\nu)$ is the difference between frequency-dependent instrumental
phases between stations, for example, imparted by the analog chain,
which is difficult to calibrate for a single station. Here we omit instrumental
calibration factors affecting the amplitude (flux), since we will use
exclusively phase information for localization, and amplitudes can, in
principle, be
calibrated station by station (before performing VLBI).

We assume that
$\phi_{ab}(\nu)$ is not direction dependent, meaning the contribution from
differences in the beams is small. In practice, beam phase constitutes a
significant systematic error that, in the KKO system, has been found to be as
high as a radian at the periphery of our field of view.
As we will show below, this is nonetheless tolerable for
our scientific objectives and we leave careful consideration of this error
to future work.

It is often useful to phase reference a visibility to an observation of a
calibrator $c$, which cancels out much of the unknown systematic contributions
to the phase. After phase referencing,
the visibility signal is given by
\begin{align}
    V^s_{ab}(\nu, t_s){V^c_{ab}}^*(\nu, t_c) 
    = S^s_\nu S^c_\nu \exp\{
        &i2\pi \nu [
            \tau_{ab}( \hat n^s, t_s) - \tau_{ab}( \hat n^c, t_c)
            + \tau^\text{cl}_{ab}(t_s) - \tau^\text{cl}_{ab}(t_c)
        ]
        \nonumber\\
        & +i2\pi \kappa [
            \text{sTEC}_{ab}^s(t_s) - \text{sTEC}_{ab}^c(t_c)] /\nu
      \}.
    \label{e:vis_cal}
\end{align}
This is the data product from which we will infer the localization.
As written, the time
dependence has been kept general, however, we will generally assume
sufficiently brief observations such that the target and calibrator
observations can be integrated down a visibility spectrum at a single effective
time. That is, $t_s$ and $t_c$ are single numbers, whereas the frequency $\nu$
covers the full observing band. There are subtleties in removing the most
rapidly changing contributions to the phase prior to integrating that are
discussed in detail by \citet{2024arXiv240305631L}. In addition, for
pulsars and FRBs, the dispersion delay in time of arrival makes the observation
time dependent on frequency, which we discuss briefly in
Appendix~\ref{a:disp_fringe}. Both these effects are critical to account for
when analyzing the data, but are not essential for motivating calibration
strategies. As such, we will not discuss them further here.

Note also that for such short (effectively instantaneous) observations where
Earth rotation is negligible, the dependence of the visibility
on the source location is through a single intermediary variable,
$\tau_{ab}(\hat n^s)$. As such, only one component of $\hat n^s$ can be
measured, the one colinear with the baseline. Two non-colinear baselines are
required to precisely measure all components of the localization.

The key aspect of Equation~\ref{e:vis_cal} is its dependence on the source location $\hat
n^s$, through the delay model $\tau_{ab}(\hat n, t)$.
Before discussing the other contributions to the phase, it is instructive to
consider how sensitive this term is to the localization so that we can
develop a rough understanding of how well other terms will need to be known.
While the delay model includes many effects, including relativistic delays and
contributions from the refractive index of the troposphere, the largest contribution is purely
geometric, such that $\tau_{ab} \approx \vec b_{ab} \cdot \hat n / c$, where $\vec
b_{ab}$ is the baseline vector connecting telescopes $a$ and $b$. This quantity
is of order $b_{ab}~\theta / c$, where $\theta$ is sky angle from zenith.
Thus, constraints on the delay at precision $\sigma_\tau$ fix the source
location to a small range of angles of width
$\sigma_\theta \sim \sigma_\tau c / b_{ab}$.
For our localization requirement of
$\sigma_\theta \approx 50\,\textrm{mas}$ (as motivated in Section~\ref{s:objectives}) and baselines
$b_{ab} \sim 1000\,\textrm{km}$, we have
$\sigma_\tau \sim 0.8\,\textrm{ns}$. As such, we must be able to isolate the
$\tau_{ab}( \hat n^s, t_s)$ from the other terms (through calibration or
otherwise) to nanosecond 
precision, or $\sigma_\phi = 2 \pi \sigma_\tau \nu \sim 3\,\textrm{rad}$ when
converted to phase units at 600\,MHz. This justifies our above decision to defer
consideration of beam phase, which in the worst cases
contributed $\sim 2\,\mathrm{rad}$.

In practice, the dependence of the visibility on the target localization is
best viewed as coming through the combined quantity
$\tau_{ab}( \hat n^s, t_s) - \tau_{ab}( \hat n^c, t_c)$, which we henceforth
refer to as the geometric delay. This makes it clear that
the calibrator position $\hat n^c$ must be known to better than the
localization requirement. It also hints at reduced sensitivity to errors in the
delay model (including uncertainties in the station locations) when the
calibrator is near on the sky to the target. However, in practice our delay
model \texttt{calc11} \citep{Eubanks91} has been validated to precision exceeding our needs, and
the baselines can be calibrated as shown by \citet{2024AJ....168...87L}.

The geometric delay must be separated from other contributions to the
interferometric phase. The first of these is
$\tau^\text{cl}_{ab}(t_s) - \tau^\text{cl}_{ab}(t_c)$,
describing the time-transfer of our clocking
solutions. In the previous section and in \citet{2022AJ....163...48M}, we described
hardware solutions that render this term negligible compared to our
localization specifications provided $|t_s - t_c| \lesssim 2000$\,s. 

The quantity $\text{sTEC}_{ab}^s(t_s) - \text{sTEC}_{ab}^c(t_c)$ represents the
difference of the slant TECs from the ionosphere between the two lines of sight
and between the target and calibrator observation times and directions. Thus, it is
expected to be smaller for target--calibrator observations that are more
proximal in time and sky location. We see from Equation~\ref{e:vis_cal} 
that the geometric delay causes a
linear phase in the visibilities with frequency
whereas the dispersive delays induce phase that goes as $1/\nu$. As such, we
anticipate that given a sufficiently wide band and \SNR, these contributions
should be non-degenerate and no further ionospheric calibration should be
necessary. However, if statistical power is limited, degeneracy between the
geometric and dispersive delays can substantially reduce localization
precision. A prior on the differential slant TEC can help to break the degeneracy,
motivating more proximal calibrators to enable
tighter priors (for further discussion, see Appendix~\ref{a:tec_prior}).

The phase-referencing and localization fitting procedure described above
motivates the desire for calibrators and observations thereof with the following properties:
\begin{enumerate}
    \item Compact on our $\sim3000$\,km baselines within our 400--800\,MHz observing band.
    \item Known astrometric precisions substantially better than our
        localization precision target of 50\,mas.
    \item Sufficient \SNR{} observations such that the noise in
        the phase-referenced visibilities $V^s_{ab}{V^c_{ab}}^*$ observation is
        not dominated by noise from the calibrator observation. Since we wish
        to localize FRBs with $\SNR_\times > 6$, calibrator observations with
        $\SNR_\times \gtrsim 20$ are desired. This can be
        achieved through a combination of calibrator brightness and observation
        duration.
    \item Calibrator observations separated in time from target observations by
        less than 2000\,s, such that time-transfer of our clocking solutions
        does not introduce large localization errors.
    \item Calibrators with Earth-frame sky locations as close as possible to
        the target FRBs such that as narrow a prior as possible can be placed
        in the slant-TEC differences. Given CHIME's drift-scan
        design and the shape of its beam, this implies a calibrator at a
        similar declination to the target that transits within the required
        time frame.
\end{enumerate}

\subsection{Potential calibrators}


\subsubsection{VLBI continuum calibrators}
\label{s:contcal}

While global VLBI solutions exist for
frequencies\footnote{\texttt{https://astrogeo.smce.nasa.gov/sol/rfc/rfc\_2024b.}}
$>1$\,GHz, many of
these sources are unusable at $\sim$600\,MHz due to low frequency turnover
(e.g.,~from synchrotron self-absorption), larger physical sizes of the emitting regions
at low frequency and hence sources being resolved, and in some cases angular scatter broadening.
Nevertheless, a survey conducted by \citet{2024arXiv240911476A}
detected over 200 calibrators on the CHIME--GBO baseline from the radio
fundamental catalog \citep{2025ApJS..276...38P} which are sufficiently bright ($\sim 500$ mJy) to be observed in standard $\sim$100-ms baseband captures. 
The density is such that, on average, there are about 2 such sources in CHIME's field of view at a time. \citet{2024arXiv240911476A} also showed that with
a deeper survey (1.4-s
integration times) our calibrator density is expected to increase by a factor
of $\sim 7$. Such a survey is currently underway.

\subsubsection{Pulsars}

Pulsars allow time-domain
separation from constant backgrounds, and the pulsed emission is guaranteed to be compact (apart from
scatter angular broadening for a small subset).
Additionally, pulsars are
sufficiently abundant that the Outriggers can observe them frequently enough to
maintain phase coherence across our network. \citet{2022AJ....163...65C,2023arXiv230709502C}
described and demonstrated an end-to-end a calibration procedure
relying on pulsars as calibrators (e.g., using giant pulses from the Crab
pulsar (PSR B0531+21) detected on the baseline between CHIME and the Algonquin Radio Observatory, located $>$\,3,000\,km from CHIME).

Independent of our ultimate calibration strategy, the fact that dispersed pulsed
emission from pulsars closely resembles FRBs makes these sources critical
checks of our localization capabilities. By analyzing pulses from pulsars as
if they were FRBs (as well as the
small number of repeating FRBs localized by other instruments), we can verify
the end-to-end performance of the Outrigger system. Such a performance
characterization strategy has already been used for KKO
\citep{2024AJ....168...87L} and CHIME/FRB's baseband system
\citep{2021ApJ...910..147M}.

\begin{figure*}
\vspace{-1mm}
\centering
\includegraphics[width=0.99\linewidth]{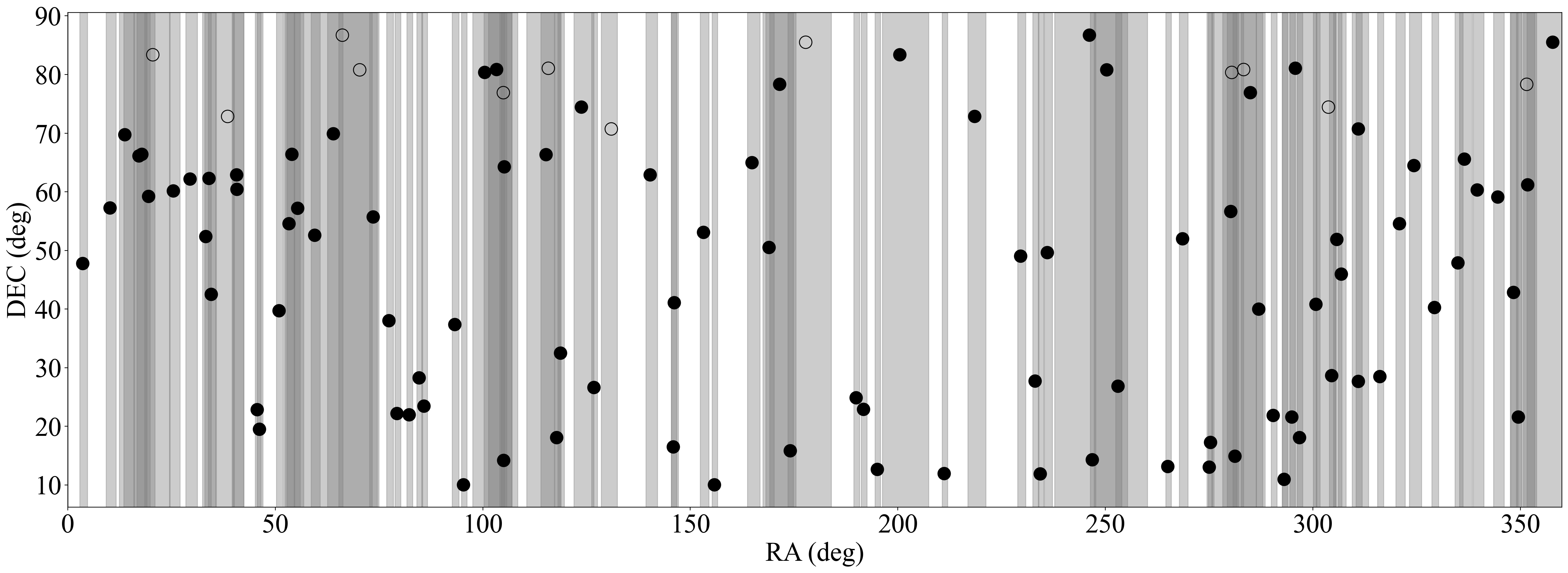} 
\caption{Cartesian projection of the 97 pulsars that the CHIME/FRB Outriggers
    will use as calibrators~\citep[][Curtin et al. 2025, in
    preparation]{Pearlman2024}.
    Pulsars are shown with black circles while the
    grey circles show the lower transits for the high-declination pulsars. The
    width of the grey shaded area indicates the range of hour angles that the
    pulsar is observable by CHIME, which depends on the declination of the
    source. Darker regions indicate that more than one pulsar
    is visible in the CHIME beam at the given time. 
    Note that CHIME only sees the sky north of declination $-11$ degrees.
    Coordinates are J2000.}
\label{fig:skyCoverage}
\end{figure*}

We have selected a sample of 97 pulsars to use as a possible calibrator sample for the CHIME/FRB Outriggers.
Each of our targets is sufficiently bright to have a phase-folded detection
significance \SNR{} of at least 15~in a single CHIME
transit\footnote{Transit time is a function of declination, with a minimum of
$\sim$10 min for Southern declinations to hours close to the North celestial
pole.}. This detection threshold is sufficient to detect the pulsars in
cross-correlation between CHIME and the Outriggers. We have prioritized pulsars
that have no measured temporal broadening, selecting those sources with implied
angular broadening less than 10\,mas at 400\,MHz. However, this criterion
creates a sample with non-ideal sky coverage. We have relaxed this parameter
for nine targets with critical hour angles. The sky distribution of the 97
calibrator pulsars is shown in Figure~\ref{fig:skyCoverage}.

The design plan for the Outriggers allows for an astrometric error of up to 50
mas, of which roughly 40 mas is budgeted for systematics and the ionospheric
uncertainty. Therefore, to stay within the remaining error margin when added in quadrature, our pulsar
calibrator positional uncertainties must be $<30$ mas. While $\sim$mas-level
localizations can sometimes be achieved for millisecond pulsars using pulsar
timing, there are currently not enough such pulsars to act as the calibrators
for the Outriggers.
Further complicating matters,
pulsars typically have proper motions of $\sim$10\,mas/year, with some displaying proper motions of
$>$\,50\,mas/year \citep[e.g.,][]{2009ApJ...698..250C, 2019ApJ...875..100D}.

Motivated by the need to have pulsar locations determined precisely and
accurately, we have observed 84 pulsars at 7 different epochs with the Very Long
Baseline Array (VLBA) since February 2021, in order to obtain precise positions
and proper motions (VLBA/21A-314, PI: Kaczmarek; VLBA/22A-345,
PI: Curtin; VLBA/23A-099, PI: Curtin; VLBA/24B-328). The remaining 13 pulsars have previous
VLBA positions and proper motions from either \citet{2009ApJ...698..250C} or \citet{2019ApJ...875..100D}.
Each pulsar is observed in the 1.35--1.61\,GHz frequency band. As each
pulsar has a well determined timing ephemeris derived from the CHIME/Pulsar
instrument, we employ matched-filter pulsar gating to increase the
\SNR{} per observation. Additionally, as absolute position
measurements are imperative, each observation is phase-referenced. So far,
we have obtained 360 hours of pulsar observations
with the VLBA\@. This program will not only result in a pulsar calibrator survey that
can be used across a variety of frequencies, but will also double the number of
pulsars with well measured positions, parallax, and proper motions. Results
from this pulsar astrometry project will be published separately (Curtin et al.
2025, in preparation).

\subsection{Observational Capabilities}
\label{s:obscap}

CHIME and the Outriggers are capable of several
observation modes, including the station $N^2$ correlation, full-array baseband
captures, and tied-array tracking beams. The first of these is the traditional
interferometric mode where each of a station's antennas is correlated against
all of the others. While critical for internally calibrating and characterizing
a station, it discards the global phase information required for VLBI\@. In
contrast, full-array baseband captures and tied-array tracking beams retain
global phase information and can be used for VLBI\@. Here we describe these two
modes in more detail.
All three of these data acquisition modes are run commensally.

Data from either the full-array baseband capture system or the tied-array tracking beams
may be correlated between stations for VLBI\@. The details of this process have
been provided by \citet{2024arXiv240305631L}.

\subsubsection{Full-array baseband acquisitions}
\label{s:baseband}
Each Outrigger is equipped with a full-array baseband capture system as
originally implemented for CHIME/FRB for more detailed characterization of
FRBs \citep{2018ApJ...863...48C, 2021ApJ...910..147M}
and first used for VLBI with
the CHIME Pathfinder \citep{2021AJ....161...81L}. For this system, CHIME and
the Outriggers store baseband data for all antennas in a ring buffer on the
X-engine nodes. Short subsets of these data may be excised from the
buffer and written to disk. Functionality exists to account for dispersion
delay, such that the time-span captured is different for each frequency
channel.

The key advantage of recording data in
full-array mode is that it allows station beams to be formed in any direction within the
200-square-degree field of view of each station in subsequent offline analysis. This
provides the ability to observe any target or calibrator within the field of
view at the time of recording. Critically, because the sky locations of FRBs
are not known a priori, this is the only system capable of
capturing phase-preserving data for FRBs.

The main limitation of the baseband capture system is that, because we record data
for all antennas in a station's array, the high data rates
limit the duration and frequency of these captures. The full-array baseband
data rate is 6.6\,Tbps for CHIME, 0.41\,Tbps for KKO, and 0.82\,Tbps for the GBO
and HCRO\@ stations. Due to limitations in transferring data off the X-engine nodes,
captures are limited in duration to the length of an intermediate 1.4\,s `readout
buffer'. Multiple simultaneous captures are
possible so long as their durations do not cumulatively surpass 1.4\,s. Upon
filling the readout buffer, it takes several minutes to transfer the data to
disk so that additional data can be collected.

Functionality exists to capture full-array baseband data synchronously at CHIME
and the Outriggers either at pre-scheduled times, or upon receipt of a trigger
from CHIME/FRB, either from an FRB or a pulsar single pulse.
For CHIME, this buffer has a duration of 33\,s and for the
Outriggers it is 38\,s.
CHIME/FRB's triggering latency from pulse arrival at 400\,MHz is about 10\,s,
with the remainder of the buffer length used for the dispersion delay between
800 and 400\,MHz.
As such, the system enables capture of the full band for sources with DM up to
$\sim1000$\,pc/cm$^3$ at CHIME and the Outriggers.

\subsubsection{Tracking Beams}
\label{s:tracking_beams}

CHIME and each of the Outrigger stations will utilize
tied-array beams to digitally track compact calibrators (e.g., continuum
sources and pulsars) that will be used for VLBI calibration~\citep{Pearlman2024}. This system was
born out of the tracking beams developed for CHIME/Pulsar
\citep{2021ApJS..255....5C}, paired with a VLBI recording backend.
Simultaneous dual-polarization beams are formed to a specified position
at each site and the
resulting phase-preserving voltage data are streamed to disk in VLBI Data
Interface Specification (VDIF) format.

Since the tied-array beams collapse the information from all of a station's
antennas into a single data stream, the data rate is far lower (only 6.4\,Mbps
per beam for each station).
This capability allows for much longer duration
observations for sources for which the sky location is known a priori, such as
pulsars and continuum calibrators.
As such, observations that span the duration of a source's transit through
CHIME's field-of-view are possible, which can be 10 minutes or longer depending
on the source's declination. We initially plan to use two such
tied-array beams simultaneously.
This system has been designed to ensure the real-time tied array beamformer and the
offline beamformer for the full-array baseband capture system have consistent
pointing and phase centers, allowing calibration solutions derived from one
to be applied to the other.

\subsection{Calibration Strategies}

Having discussed both the potential calibration sources as well as our
observational capabilities, we now sketch out a few calibration strategies that
would allow us to achieve our scientific requirements. As stated previously,
which of these strategies is used will depend on the resolution of lingering
uncertainties about calibrator abundance, the ionosphere, and
logistical considerations. However, the strategy described in
Section~\ref{s:triggered_cal} is promising in the long term and is a current
development target.

\subsubsection{Tracking rare calibrators and time transfer}
\label{s:tracking_cal}

At the beginning of the CHIME/FRB Outriggers program, the abundance of long-baseline
continuum
calibrators in the CHIME band was known but incomplete: existing catalogues
covered either similar baseline lengths or similar frequencies but not
both~\citep{2008ApJ...673...78L,2015A&A...574A..73M}. There was a risk
that our calibrator
grid would be sparse: perhaps limited to pulsars and a small fraction of the continuum
calibrators used at higher frequency.

As such, our initial calibration strategy---designed to mitigate this unlikely but potentially catastrophic
scenario---was to continuously monitor up to two
pulsars using the tracking beam capabilities described in
Section~\ref{s:tracking_beams}. Tracking beams provide ample observation
lengths to achieve high $\SNR$ on even dim sources while keeping data volumes
manageable. As can be seen from Figure~\ref{fig:skyCoverage}, there exist
$\sim$1000-s periods for which there are no pulsars in the CHIME field of
view, requiring the on-site atomic clocks described in
Section~\ref{s:station_design} to interpolate timing solutions
between transits.

While this strategy provides a means to calibrate timing offsets, it provides
little control over target--calibrator sky separations. As such, the geometric
and dispersive delays have to be disentangled by fringe-fitting, with wide
priors on the differential ionosphere. Nonetheless, in
Section~\ref{s:forecasts}, we show that it is still possible to achieve our
localization precision specifications in most cases using this strategy.
The concern over calibrator abundance has been relieved substantially by
a direct VLBI survey performed on the CHIME--GBO baseline, as described in
Section~\ref{s:contcal}. Nevertheless, the clocking and data acquisition
infrastructure already developed opens the possibility of
future ancillary-science applications
of the CHIME/FRB Outriggers.

\subsubsection{In-beam calibrators in full-array baseband dumps}
\label{s:fullarray_cal}

In the opposite extreme to the previous case, it is possible that continuum
calibrators are sufficiently abundant that there would reliably be suitable sources in
the same triggered baseband captures that contain our FRB targets. From the
full-array baseband data it is possible to phase the station arrays to any
source within the field of view, meaning that the only additional data collection
required would be to lengthen the duration of the captures to improve $\SNR$.

One key advantage of this strategy is that the calibration observation is
contemporaneous with the target observation, which completely eliminates the
need to time-transfer the calibration solution as well as any concern about
time variability of the ionosphere.
The primary drawback of this strategy is the limitations on the durations of
the baseband captures, with 1.4\,s being a hard limit which causes several
minutes of dead time in our ability to capture subsequent targets.
This severely limits the $\SNR$ that can be achieved
on the calibrators. Also problematic are data volumes, particularly for
CHIME where the 1024 feeds yield data at a rate higher than a
tied-array beam by the same factor.
Nonetheless, this strategy has
already been found to be an effective calibration strategy
on our shortest CHIME--KKO baseline, which has the weakest requirements on
calibration sources due to the reduced ionosphere and fewer sources being
resolved on the shorter baseline \citep{2025arXiv250211217A}.

\citet{2024arXiv240911476A} showed that this strategy is promising even on the
CHIME--GBO baseline, where an average of about 2 sources can be
detected in a 100-ms baseband capture. The modest source density makes this
strategy vulnerable to Poisson fluctuations in the number of sources in the field of view
during an FRB detection.
Obtaining sufficiently high $\SNR$
observations to calibrate requires lengthening the duration of the baseband
captures, and even then does not always yield a calibration solution.  This
strategy is also insufficient to achieve very small target-calibrator
separations, meaning, as in the previous strategy, fringe fitting is required to
separate the geometric and ionospheric delays.
Nonetheless, when one of the brighter sources falls within the
same field of view as a target, this strategy provides a convenient
calibration.

\subsubsection{Triggered calibrator follow-up}
\label{s:triggered_cal}
As discussed in Section~\ref{s:contcal}, we now know that on average about 14
compact continuum calibrators are present in the CHIME field-of-view at any
given time.
However, detecting most of them requires a 1.4-s long observation (the maximum for
the full-array baseband system), and obtaining the requisite
$\SNR_\times \approx 20$ for phase referencing requires substantially longer (e.g.\ 10 to 100\,s).

This motivates a calibration strategy where
we could use the low-latency FRB trigger
to also trigger longer tracking beam observations of a previously verified
continuum calibrator~\citep{Pearlman2024}. The tracking beams readily provide a longer
integration and thus ample $\SNR$.
It would be feasible to start calibrator
observations within tens of seconds of the target observation, limited
primarily by the FRB dispersion delay and triggering latency.
Using the on-site atomic
clocks for time-transfer of the calibration solution
would still be necessary, but would incur negligible
net clock error.
Furthermore, with $\sim14$ potential calibrators spread across CHIME's
120-degree long field-of-view, it would be possible to target sources
within $\sim 10$ degrees of the target, minimizing differences in the line-of-sight
sTEC\@. We show in Section~\ref{s:forecasts} that the resulting narrow prior
on the ionosphere leads to substantial gains in localization precision, even
for narrowband sources.

However, the strategy requires a pre-survey to create a catalog of confirmed suitable
calibrators to target, which, as mentioned above, is currently underway.
In addition, the logistical implementation of this strategy is
currently under development.

\section{Forecasts}
\label{s:forecasts}

In this section we perform forecasts for the localization precision of the
Outriggers for varying properties of target FRBs (\SNR, band
occupancy) as well as assumptions about the ionospheric TEC and our ability to
remove its contribution through calibration.

For our forecasts, we calculate the full posterior distribution for
fringe-fitting to Equation~\ref{e:vis_cal}. Free parameters are the RA and Dec
of the FRB ($\hat n^s$), with independent nuisance parameters for each of our
three CHIME--Outrigger baselines for the differential clock delay
($\tau^\text{cl}_{ab} - \tau^\text{cl}_{ab}$)
and the
differential sTEC ($\text{sTEC}_{ab}^s - \text{sTEC}_{ab}^c$). We place
priors on these nuisance parameters, as described in the following paragraphs.
Details of the
implementation are provided in Appendix~\ref{a:forecast}.

To place a prior on the differential sTEC nuisance parameters, we build a model
for the distribution of these parameters by randomly sampling target and
calibrator lines of sight from the International Reference Ionosphere
\citep[IRI,][]{ISO16457}
model. Details of this procedure are provided in Appendix~\ref{a:tec_prior},
and we denote this our \emph{fiducial} model for the ionospheric priors.
Sampling target and calibrator locations randomly from the field-of-view mimics
calibration schemes where sources are rare, such as those relying on pulsars
(Section~\ref{s:tracking_cal}) or, on our longer baselines, when integration time is
limited to the duration of full-array captures
(Section~\ref{s:fullarray_cal}).

In addition to the fiducial prior, we also consider two other cases. First, we envisage
developing a calibration strategy or ionospheric measurement or modelling
procedure that enables us to null or subtract 80\% of the
ionosphere on all baselines. For example, the strategy described in
Section~\ref{s:triggered_cal} could achieve target--calibrator separations of
$\sim 10$ degrees, which would null a large fraction of the ionospheric
effects.
In this scenario, we make the ionospheric priors 20\% as wide as fiducial.
Second, we consider the case where the ionosphere is
dramatically stronger than in the IRI model, where we make our ionospheric
priors a factor of 5 wider than in the fiducial case. Such an extreme case
is unlikely, but can be viewed both as a worst-case scenario and as
illustrative of the ability to fringe-fit for the ionosphere even without external
constraints.
These models define the priors $p(\text{sTEC}_{a})$
used in Appendix~\ref{a:forecast}.

We also require a prior on the differential clock error on each baseline. For
the CHIME--KKO baseline, where we have already determined that there is a high
density of in-beam calibrators, we assume this error to be negligible. For the
CHIME--GBO baseline, both sites have a hydrogen maser clock, and so we assume a
Gaussian distributed prior on the clock offsets with a standard deviation of
100\,ps. For CHIME--HCRO, where one site has only a rubidium clock, we assume
that the clock offsets have a standard deviation of 200\,ps.

Finally, there will unavoidably be additional systematic error. For example,
differential beam phase could exist between the two stations. This has yet to be
studied in detail, although there have been isolated rare cases of unexplained
phase residuals of $\sim1$\,radian on the CHIME--KKO baseline at the top of
the observing band. For simplicity, we model such systematics as an additional
delay error with a standard deviation of 300\,ps (corresponding to 1.5\,radians
at the top of the band), which is added in quadrature to the clock offsets
described above. This, combined with the previous paragraph defines the
Gaussian prior $p(\tau^\text{cl}_{a})$ used in Appendix~\ref{a:forecast}.

We consider FRBs that occupy 400, 200, and 100~MHz effective bandwidth, all
assumed to be centred at CHIME's band center of 600\,MHz.
We parameterize the FRB brightness by $\SNR_\textrm{CHIME}$, the
signal-to-noise ratio as seen by CHIME alone. We assume that the cross-correlation
$\SNR_\times$ is $\SNR_\textrm{CHIME}/\sqrt{8}$ for the CHIME--KKO baselines and
$\SNR_\textrm{CHIME}/\sqrt{4}$ for the CHIME--GBO and CHIME--HCRO
baselines (from Equation~\ref{e:snr_times}). These assumptions neglect the fact
that each site has RFI bands that
do not overlap perfectly (we defer study of this effect to future work).

For simplicity, we assume that on each baseline, a calibrator with
$\SNR_{\times,c}=20$ is observed. Note that this implies that a net brighter
calibrator is observed for the less-sensitive CHIME--KKO baseline, however,
calibrators have been found to be abundant on the short baseline. As such, for
the phase-referenced visibilities
$(\SNR_{\times,\text{sc}})^{-2} = (\SNR_{\times,\text{s}})^{-2} +
(\SNR_{\times,\text{c}})^{-2}$.

We assume that the \SNR{} is evenly distributed
across the full band occupied by the FRB\@. Targets are assumed to be observed at
CHIME's zenith, which affects the baseline projections but not signal strength
since we are parameterizing FRBs using $\SNR_\textrm{CHIME}$.

\begin{figure}
    \centering
    \includegraphics[width=0.98\linewidth]{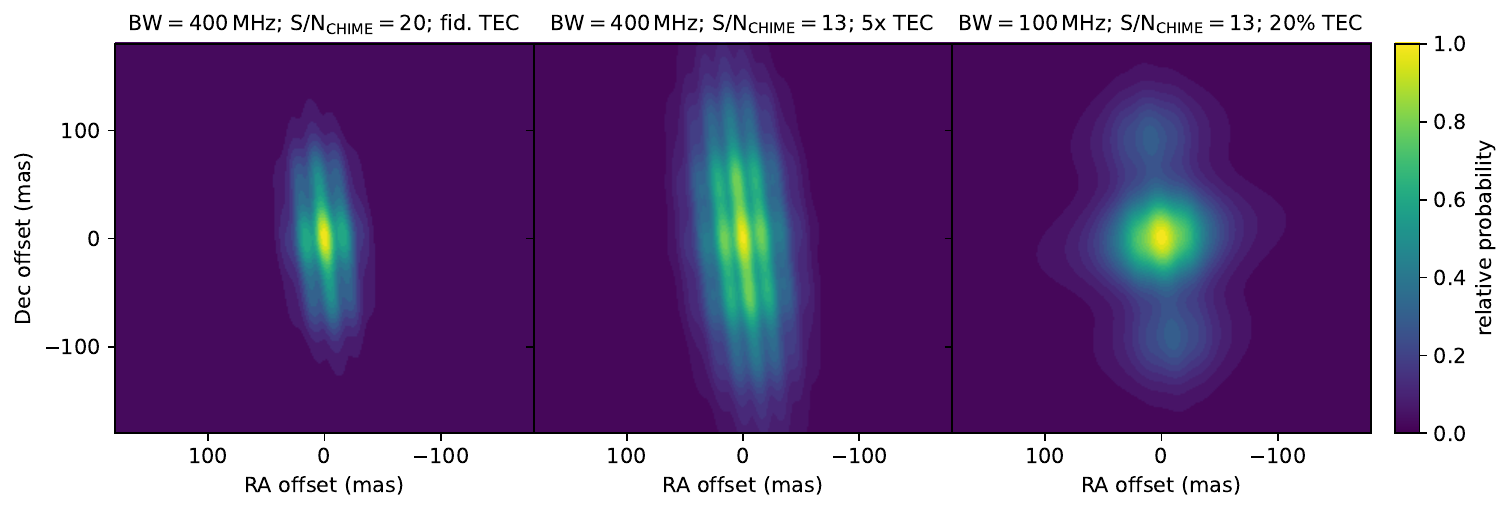}
    \caption{Forecasted localization region for FRBs with three combinations of bandwidth occupancy,
    signal-to-noise ratio, and assumptions about the impact of the ionosphere.
    In each case parameters are listed above the figure. \textsl{Left:} we forecast that we
    achieve our localization precision goal of 50\,mas for fairly typical FRBs
    with $\SNR_\textrm{CHIME}=20$ filling the observing band and fringe-fitting for
    the differential sTEC in our fiducial ionosphere model. \textsl{Centre:}
    even at lower $\SNR_\textrm{CHIME}$ and an ionosphere five times thicker than our
    fiducial model, the wideband fringe fits yields reasonably good
    localizations. \textsl{Right:} for a narrowband FRB, precise localization requires
    a calibration that
    nulls 80\% of the ionosphere.
    \label{f:localizations}
    }
\end{figure}

Full localization posterior distributions are shown in
Figure~\ref{f:localizations} for three permutations of the band occupancies,
 $\SNR$s, and ionospheric scenarios considered. Several features are apparent:
the localization ellipses have roughly the 3 to 1 aspect ratio expected from
the ratio of baseline lengths involving the GBO and HCRO stations (whose information
dominate over that from KKO). The orientation is as expected
for GBO's location to the East and South of CHIME\@. When the full bandwidth is
available, distinct lobes are visible and are associated with local likelihood maxima in the
fringe fit for delay and sTEC\@. These are less pronounced in narrower-band
observations. The overall envelopes of the localizations are set by a
combination of the width of the ionospheric prior and \SNR{}.

\begin{figure}
    \centering
    \includegraphics[width=0.90\linewidth]{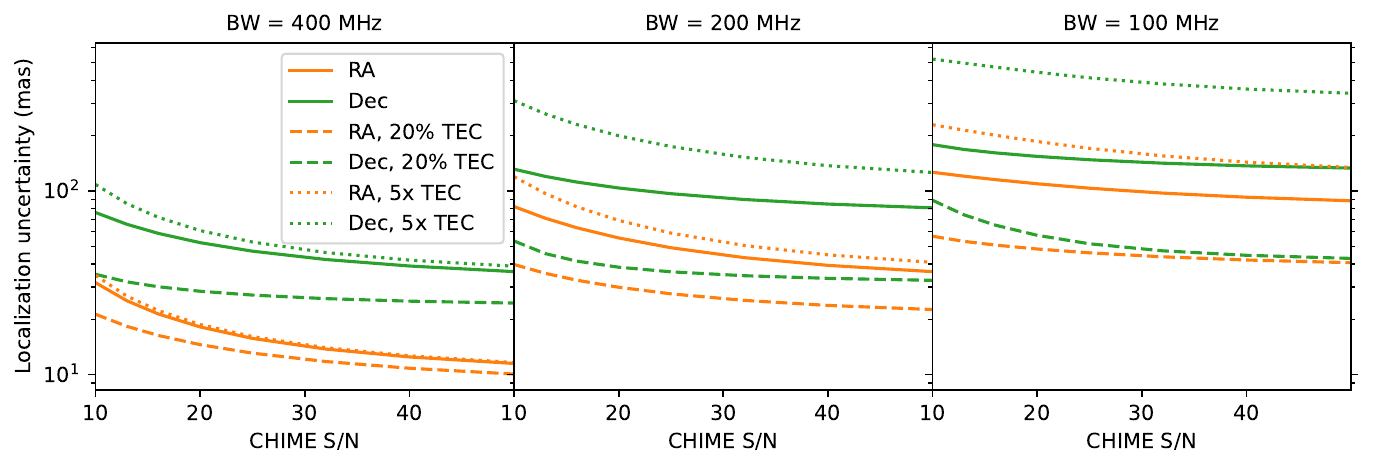}
    \caption{%
    Localization forecasts for the CHIME/FRB Outriggers. Forecasts are
    provided as a function of FRB band occupancy (see text above the left,
    centre, and right panels), and signal-to-noise ratio as observed by CHIME\@.
    We show three different scenarios for the ionosphere: 1.~our fiducial
    ionospheric prior model
    where we must fringe-fit for the differential sTEC between targets and
    calibrators randomly drawn from the field of view; 2.~a scenario where we
    find nearby calibrators that allow us to null 80\% of the ionospheric
    contribution; and 3.~a scenario where the ionosphere is 5 times thicker than in
    our fiducial model widening the prior distribution by a factor of 5.
    We forecast that we will achieve our localization goal
    of $\sim$50\,mas precision in both dimensions for all band-filling FRBs
    except under the most pessimistic assumptions, and that even narrowband
    FRBs can be localized provided the majority of the ionosphere can be
    nulled.
    \label{f:forecasts}}
\end{figure}

As summary statistics for the localization uncertainty, we use the standard
deviations of the RA and Dec, which, given our array geometry, are only slightly
misaligned with the major and minor axes of the localization envelope.
This is plotted for all FRB properties and
ionospheric scenarios considered in Figure~\ref{f:forecasts}.
Note that in our model, the distributions of the data errors and nuisance
parameters are all symmetric and centered at zero, so there is no net bias in
the localizations.

We see that for
band-filling FRBs,
the fringe fit is able to distinguish between the
non-dispersive delays that contain localization information and dispersive
delays from the ionosphere, roughly reaching our $\sim$50\,mas precision target
even for modest \SNR{} and very pessimistic assumptions about the
ionosphere. 
We note that band-filling FRBs represent 85\% of CHIME/FRB sources \citep{2025ApJ...979..160S},
and that narrow-band FRBs are preferentially repeaters, providing multiple opportunities to
localize \citep{2025ApJ...979L..21S}.

In all cases, our localization in the RA direction is far less sensitive to
ionospheric assumptions than the Dec direction. This is likely because the
CHIME--KKO baseline, which is predominantly EW but
short enough that it experiences negligible differential sTEC, provides a prior
on the RA localization that is mostly independent of the ionosphere. This breaks the
degeneracy between dispersive and non-dispersive delays in the CHIME--GBO
fringe fit, i.e., the baseline picks a lobe of an otherwise degenerate
localization. This is more effective in wider band observations where these
lobes are more pronounced.

Developing a calibration strategy that removes the
majority of the ionospheric contribution (such as that
described in Section~\ref{s:triggered_cal}) would allow us to significantly exceed
our precision targets. The 20\% TEC case in Figure~\ref{f:forecasts} shows that in this scenario,
localization precision
quickly hits its systematic-error floor even at modest \SNR{}, presenting
opportunities to achieve even higher precision should those systematics (clock
error and beam phase) be mitigated.

\section{Current Status and Outlook}
\label{s:status}
At the time of writing (early 2025), the CHIME/FRB Outriggers project has made
significant strides towards achieving its scientific goals. Two of the three
Outriggers, KKO and the GBO station, are fully constructed and operational, while
the HCRO station has achieved first light and is under commissioning.

Construction of the KKO station was completed in the Summer of 2022. Since then, it
has been collecting valuable data, including baseband captures of FRBs. The
commissioning phase, which extended through to the Summer of 2023, demonstrated
the station's instrument performance and scientific capabilities as shown by
\citet{2024AJ....168...87L}. Follow-up and analysis of the initial data from KKO
are ongoing, with early results
indicating promising avenues for further research \citep{2024ATel16682....1L, 2025ApJ...979L..21S,
 2025ApJ...979L..22E, 2024arXiv240911476A, 2025arXiv250211217A}.

Construction of the GBO station was completed in Spring of 2023. Since becoming
operational, it too has consistently been gathering data.
The GBO station is now operating reliably, successfully detecting and localizing both pulsars and FRBs.

After an extensive local permitting process, the HCRO station saw the
completion of its telescope structure and reflector in
the Summer of 2024, which
was followed by deployment of the analog instrumentation.
Deployment of the 
digital instrumentation occurred in early
2025, with first light achieved shortly thereafter. The station is now in
full commissioning. Once fully
operational, the HCRO station will complement the other two stations,
providing
two-dimensional localizations.

Tests that verify KKO's performance were developed by
\citet{2024AJ....168...87L}. Replicating these for the full network is
currently underway, and will be a key step toward transitioning to science
operations. Chief among these tests is the localization of roughly a hundred
single pulses from about 20 distinct pulsars as if they were FRBs. Comparing
these to externally measured pulsar positions provides a means to characterize
the end-to-end localization performance. This will in turn provide confidence
in the robustness of our localizations and their uncertainties.

In terms of the observational capabilities described in Section~\ref{s:obscap},
both the $N^2$ system and the
full array baseband captures have been deployed and are in use. The tied-array
beams are still in active development. Our forecasts in Section~\ref{s:forecasts}
have shown that nulling as much of the ionosphere as possible is critical for
localization precision, especially for narrowband FRBs. As such, the deployment
of observational capabilities to access fainter calibrators that are
closer to target FRBs will provide a substantial boost to our localization
precision.

Looking ahead, the CHIME/FRB Outriggers project is set to make substantial
contributions to the field of FRB research. With the completion of all
three stations, the project will localize a large fraction of CHIME-detected
FRBs,
enabling both detailed studies of rare sources and the statistical analysis of FRB
properties and their host environments. Continued efforts to refine calibration
techniques will further improve localization accuracy, addressing challenges
posed by ionospheric variations and other systematic errors.
With the bulk
commissioning having occurred near solar maximum, and upcoming operations to
take place during a period of declining solar activity, dealing with the
ionosphere (one our biggest challenges) will become progressively easier.
As such, we expect the performance of the Outriggers to continue to improve
over the coming months and years.

The data collected by the CHIME/FRB Outriggers will be instrumental in
advancing our understanding of the origins, environments, and characteristics
of FRBs. Collaborative efforts with other observatories and research
initiatives will amplify the scientific impact of the project. Overall, the
CHIME/FRB Outriggers project is poised to become a leading contributor to FRB
research, providing valuable insights and paving the way for future discoveries
in astrophysics.

\section{Acknowledgments}

We acknowledge that CHIME and the \kkoname{} Outrigger (KKO) are built on the
traditional, ancestral, and unceded territory of the Syilx Okanagan people.
K'ni\textipa{P}atn k'l$\left._\mathrm{\smile}\right.$stk'masqt is situated on
land leased from the Imperial Metals Corporation. We are grateful to the staff
of the Dominion Radio Astrophysical Observatory, which is operated by the
National Research Council of Canada. CHIME operations are funded by a grant
from the NSERC Alliance Program and by support from McGill University,
University of British Columbia, and University of Toronto. CHIME/FRB Outriggers
are funded by a grant from the Gordon \& Betty Moore Foundation. We are grateful
to Robert Kirshner for early support and encouragement of the CHIME/FRB
Outriggers Project, and to Dusan Pejakovic of the Moore Foundation for
continued support. CHIME was funded by a grant from the Canada Foundation for
Innovation (CFI) 2012 Leading Edge Fund (Project 31170) and by contributions
from the provinces of British Columbia, Québec and Ontario. The CHIME/FRB
Project was funded by a grant from the CFI 2015 Innovation Fund (Project 33213)
and by contributions from the provinces of British Columbia and Québec, and by
the Dunlap Institute for Astronomy and Astrophysics at the University of
Toronto. Additional support was provided by the Canadian Institute for Advanced
Research (CIFAR), the Trottier Space Institute at McGill University, and the
University of British Columbia. The CHIME/FRB baseband recording system is
funded in part by a CFI John R. Evans Leaders Fund award to IHS.

\allacks

\bibliography{refs,frbrefs,psrrefs}

\appendix

\section{Effect of frequency-dependent observation time for dispersed pulses}
\label{a:disp_fringe}

We consider the primary effect of Earth rotation during the
observation, showing that the additional information therein is hard to exploit, and
therefore
solidifying the need for at least two baselines.

Due to the substantial dispersive time-sweep of FRBs, one can ask
whether Earth rotation aperture synthesis can provide any 2D information
on a single baseline. To consider this, take the observation time to be the
dispersed time-of-arrival of the FRB:
\begin{equation}
    t = t_\infty + k_\text{DM} \text{DM}/\nu^2,
\end{equation}
where $t_\infty$ is the pulse time of arrival at infinite frequency and
$k_{\rm DM} = (2.41\times10^{-4})^{-1}$\,s\,MHz$^2$\,pc$^{-1}$\,cm$^3$.
Then, Taylor expanding our delay model we have
\begin{equation}
    \tau_{ab}(\hat n^s, \nu) \approx \tau_{ab}(\hat n^s, t_\infty) +
     \frac{k_\text{DM} \text{DM}}{\nu^2} \frac{\partial\tau_{ab}}{\partial
     t}(\hat n^s, t_\infty).
\end{equation}
The first term above contains the vast majority of the delay and is primarily
affected by the projection of the baseline onto the source directions $\vec
b_{ab} \cdot \hat n^s / c$. The second term is affected by the small rotation of the
baseline over the FRB's dispersive delay, and is roughly proportional to
$(\vec \omega \times \vec b_{ab})\cdot \hat n^s$, where $\vec \omega$ is
Earth's angular velocity vector, and is thus sensitive to the component
of $n^s$ orthogonal to the baseline. Unfortunately, this term has the same
frequency dependence as the ionospheric slant TEC terms in
Equation~\ref{e:vis_cal}. As such, this information is degenerate in the fit
with the TEC nuisance parameter. The priors on this nuisance parameter from
globally available TEC maps are not sufficiently precise to place any meaningful
constraints on the orthogonal component
of the baseline.

\section{Fringe-Fit Forecast Formalism}
\label{a:forecast}

Here we provide details of our localization forecast procedure.
We start by simplifying our notation from Equation~\ref{e:vis_cal} with the following
assumptions:
\begin{enumerate}
    \item We consider only baselines containing the CHIME core array, such that
        the baseline can be enumerated with a single index $a$.
    \item All observations are phase referenced to a calibration source
        observation such that the geometric delays,
        clock offsets, and slant TECs are understood to be differences between
        the target and calibrator observation.
    \item We suppress all time dependence in the notation, as we treat all
        observations, of both target and calibrator, as instantaneous
        (although not necessarily simultaneous). In practice, the FRB
        observations are spread over the dispersion delay of several seconds
        and the calibrator observation may be seconds in duration. However,
        clock error and ionospheric variability on second timescales is assumed
        to be negligible.
    \item We assume that the uncertainty on the phase-referenced visibilties are
        Gaussian distributed (instead of the more accurate Chi-squared), with
        the signal to noise ratio equally distributed over the band.
    \item The data are converted to units such that the amplitude is unity and
        the uncertainty is $\sigma_{a\nu}^2 = N_{\{\nu\}} / (\SNR_a)^2$ where
        $N_{\{\nu\}}$ is
        the number of frequency bins.
\end{enumerate}

With these changes to notation, our signal model becomes:
\begin{equation}
    V_a = \exp\{
        i2\pi \nu [
            \tau_{a}( \hat n^s)
            + \tau^\text{cl}_{a}
            + \kappa \text{sTEC}_{a} /\nu^2
        ]
      \}
\end{equation}
and our data model is:
\begin{align}
    d_a(\nu) &= V^T_a(\nu) + \epsilon_a(\nu)\\
    \langle\epsilon_a(\nu)\rangle &= 0\\
    \langle\epsilon_a(\nu)\epsilon_a(\nu)^*\rangle &= \sigma_{a\nu}^2,
\end{align}
where $V^T_a(\nu)$ is the signal model evaluated at the true values of the
parameters and $\epsilon_a$ denotes the Gaussian noise on the visibilities.

The fringe-fit localization problem is a matter of fitting our signal model to
our data for parameters $\hat n^s$ while marginalizing over nuisance
parameters $\tau^\text{cl}_{a}$ and $\text{sTEC}_{a}$ for which we will
typically have some prior (see e.g. Appendix~\ref{a:tec_prior}). The challenge is the fairly high-dimensional
parameter space (2 localization parameters plus 2 nuisance parameters per
baseline, for 8 total parameters) coupled with the poorly behaved likelihood
due to the periodic nature of the signal model, yielding a multimodal parameter
space. However, we will show that marginalizing over the nuisance parameters
can be reduced to a series of size 1D integrals, which vastly simplifies the
calculation.

The log-likelihood can be to related to the familiar $\chi^2$ statistic
\begin{equation}
    -2\log\mathcal{L}(\hat n^s,\{\lambda_\mu\}) \propto \chi^2(\hat
    n^s,\{\lambda_\mu\}) = \sum_{a\nu} \frac{|d_a(\nu) -
    V_a(\nu,\hat n^s, \{\lambda_\mu\})|^2}{\sigma_{a\nu}^2},
\end{equation}
where $\{\lambda_\mu\}$ represents the set of all nuisance parameters, and the
sum runs over all baselines and frequencies.
With some manipulation this becomes (dropping any terms that do not depend on the parameters)
\begin{align}
    \chi^2(\hat n^s,\{\lambda_\mu\}) &= -2\sum_{a\nu} \frac{\Re\{d_a(\nu)V_a(\nu)^*\}}{\sigma_{a\nu}^2}\\
            \label{e:chi2}
            &= -2\sum_{a\nu} \frac{1}{\sigma_{a\nu}^2}
                \Re\{
                    d_a(\nu) \exp[
                        -i2\pi \nu (
                            \tau_{a}( \hat n^s)
                            + \tau^\text{cl}_{a}
                            + \kappa \text{sTEC}_{a} /\nu^2
                            )
                        ]
                    \}.
\end{align}

Assuming the nuisance parameters to be independent, the posterior for the full parameter space is
\begin{equation}
    p(\hat n^s, \{\lambda_\mu\} | \{d_{a}(\nu)\}) \propto
    \exp\left(-\frac{\chi^2(\hat n^s,\{\lambda_\mu\})}{2}\right) p(\hat n^s) \prod_\mu p(\lambda_\mu).
\end{equation}

The final marginalized posterior we would like to calculate is
\begin{equation}
    p(\hat n^s | \{d_{a}(\nu)\}) \propto \int
    \exp\left[-\frac{\chi^chi2(\hat n^s, \{\lambda_\mu\})}{2}\right]p(\hat n^s)
    \prod_\mu p(\lambda_\mu) d\lambda_\mu.
\end{equation}

We see from Equation~\ref{e:chi2} that each baseline has its own nuisance
parameters such that they can
nearly be thought of as separate fitting problems, with the only coupling
between baselines through $\tau_{a}( \hat n^s)$. To break the problem apart we
define the baseline-by-baseline $\chi^2_a$ and intermediate parameters
$\tilde{\tau}_a$:
\begin{equation}
    \chi^2_a =
           -2\sum_{\nu} \frac{1}{\sigma_{a\nu}^2}
                \Re\{
                    d_a(\nu) \exp[
                        -i2\pi \nu (
                            \tilde{\tau}_a
                            + \tau^\text{cl}_{a}
                            + \kappa \text{sTEC}_{a} /\nu^2
                            )
                        ]
                    \}.
\end{equation}
With this definition, the posterior for the intermediate parameters is:
\begin{align}
     p(\tilde{\tau}_a | d_{a}(\nu)) 
        &\propto
            \int d\tau^\text{cl}_{a} d(\text{sTEC}_{a})
            \exp\left[-\frac{\chi^2_a(\tilde{\tau}_a, \tau^\text{cl}_{a},
                \text{sTEC}_{a})}{2}\right]
            p(\tau^\text{cl}_{a})p(\text{sTEC}_{a})
        \\&\propto
            \int d\tau^\text{cl}_{a} p(\tau^\text{cl}_{a})
            \int d(\text{sTEC}_{a})
            \exp\left[-\frac{\chi^2_a(\tilde{\tau}_a + \tau^\text{cl}_{a},
                \text{sTEC}_{a})}{2}\right]
            p(\text{sTEC}_{a}).
\end{align}
In the second line (in a slight abuse of functional notation)
we have written $\chi^2_a$ as a function of
$\tilde{\tau}_a + \tau^\text{cl}_{a}$ to indicate that it only depends on the
sum of the two delays. After some rearrangement, this makes
it clear that the outer integral is a convolution, which can be
computed efficiently using Fourier methods.

The full posterior is then
\begin{align}
    p(\hat n^s | \{d_{a}(\nu)\}) 
        &\propto \prod_a
            \int d\tilde{\tau}_a
            p[\tilde{\tau}_a | d_{a}(\nu)]
            \delta[\tilde{\tau}_a - \tau_{a}( \hat n^s)]\\
         &\propto \prod_a p[\tilde{\tau}_a = \tau_{a}( \hat n^s) | d_{a}(\nu)]
\end{align}

With the posterior specified, what remains is to convert it to a forecast on
parameters $\hat n^s$. One possible method is to construct an estimator for the
parameters (eg.\ maximum of the posterior or its expectation value) and draw
many realizations of the data errors ($\epsilon_a(\nu)$) and nuisance parameters
($\tau^\text{cl}_{a}$ and $\text{sTEC}_{a}$) from their expected distributions
to see how the estimator
is distributed via Monte Carlo methods. Another approach is to fix the data
errors and nuisance parameter values and use the shape of the posterior
itself to represent the uncertainty.
Via Bayes' theorem, these should be roughly equivalent and we choose to do the
latter. Because we don't expect the shape of the posterior to depend strongly
on the exact realization, we choose
$\epsilon_a(\nu)=\tau^\text{cl}_{a}=\text{sTEC}_{a}=0$, which centers our
posterior on the true value of $\hat n^s$.

\section{Model for the differential slant TEC prior}
\label{a:tec_prior}

\begin{figure}
    \centering
    \includegraphics[width=0.32\linewidth,trim={0 0 7ex 0},clip]{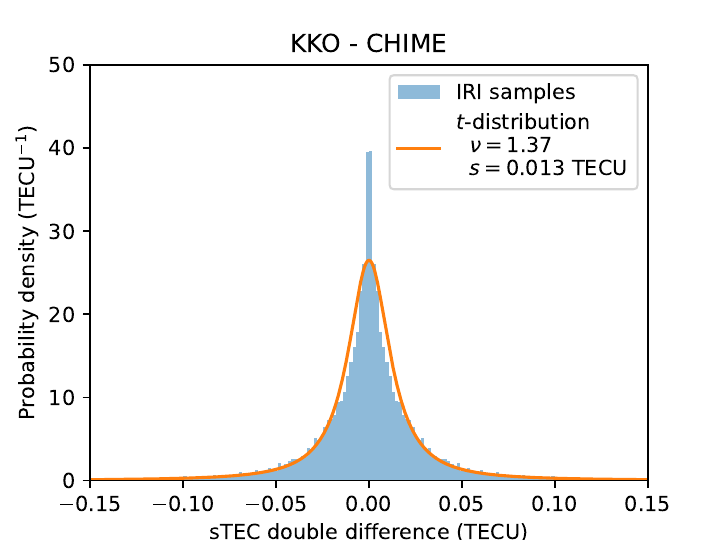}
    \includegraphics[width=0.32\linewidth,trim={0 0 7ex 0},clip]{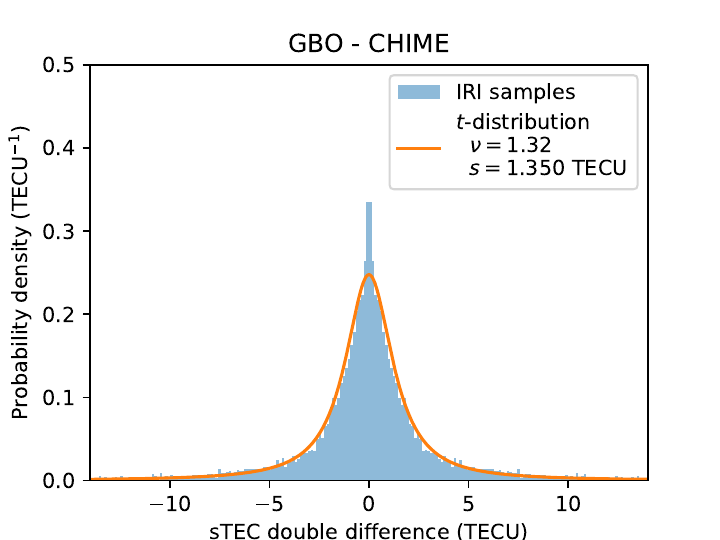}
    \includegraphics[width=0.32\linewidth,trim={0 0 7ex 0},clip]{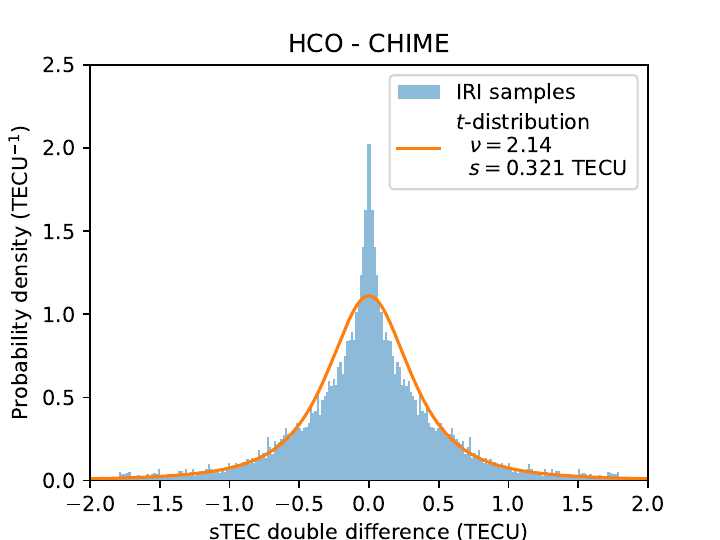}
    \caption{Samples from the IRI used to build our sTEC priors. We sample the
    sTEC at each site toward random sky locations for targets and calibrators
    within the CHIME
    field-of-view and at random times. We then calculated the
    double-difference sTEC between the target and the calibrator and between the two
    telescope sites for each baseline. The figure shows histograms of these
    differential sTEC values for each
    baseline, as  specified above each of
    the three panels. Finally, the fitted student's-$t$ distribution that we
    use for the sTEC priors in our forecasts is over plotted, with the fit
    parameters listed in the legend.
    \label{f:stec_priors}
    }
\end{figure}

We build a simple statistical model for the ionosphere based on the
International Reference Ionosphere \citep[IRI,][]{ISO16457} model, which we then use as a prior on
the sTEC nuisance parameters. Note that we are not assuming that
the IRI accurately predicts the sTEC as seen by the Outriggers, only that
sampling from it yields similar statistics as the real ionosphere. We will also
consider scenarios that are both more optimistic and pessimistic by making the sTEC prior distributions narrower and wider when determining the localization uncertainty. Our procedure
for building the model is as follows:
\begin{enumerate}
        \item We draw times randomly between Jan~1, 2023 and Jan~1, 2024.
        \item At each time, we randomly select a target and calibrator sky
            location from the CHIME meridian within 60 degrees of
            CHIME zenith, roughly mimicking CHIME's field of view.
        \item We evaluate the IRI model to determine the slant TEC toward the
            target and the calibrator at CHIME and each Outrigger.
        \item We calculate ``double difference'' slant TEC ($\text{sTEC}_{ab}^s -
            \text{sTEC}_{ab}^c$) between each
            Outrigger and CHIME and between the target and calibrator.
        \item We draw many samples, and for each Outrigger fit a distribution
            to the double difference slant TEC\@. We find that student's-$t$
            distribution (which compared to a Gaussian has much higher
            kurtosis) provides a reasonable fit to the distribution after
            fitting for the degrees-of-freedom parameter $\nu$ and scale
            parameter $s$. The
            samples and fits are shown in Figure~\ref{f:stec_priors}. 
        \item The fitted distributions are used as a baseline-dependent
            prior on the differential slant TEC nuisance parameters.
\end{enumerate}
We note that the sampled dates are close to solar maximum, whereas the
bulk of Outrigger operations will occur during a period of declining solar
activity, building some conservatism into this model.

Note that in this procedure, the calibrator and target lines-of-sight are
sampled at the same time, whereas in some calibration procedures there could be
a significant time separation, over which the ionosphere might evolve.
However, we expect that the large span of angular
separations between the target and the calibrator is the dominant contribution to the
differential sTEC, and that this
procedure generates a fairly representative distribution.

In our forecasts, we also consider variations on this model, where the
distributions are assumed to be either 5 times wider or 20\% as wide. The
former is deliberately extreme and has little physical basis. The latter,
narrower prior, is intended to roughly approximate the case where calibrators
are much more abundant. Instead of having a single calibrator over the
120-degree field of view for a typical target-calibrator separation of $\sim60$~degrees, we
envisage $\lesssim10$-degree separations being achieved. While we do not know
how the differential sTEC scales with angular separation (and the IRI model does
not capture small scales), we consider it plausible that moving from $\sim60$-
to $\lesssim10$-degree separations could result in a five-fold decrease.

\end{document}

%% file: authors.tex
\author[0000-0001-6523-9029]{Mandana Amiri}
  \affiliation{Department of Physics and Astronomy, University of British Columbia, 6224 Agricultural Road, Vancouver, BC V6T 1Z1 Canada}
\author[0000-0001-5908-3152]{Bridget C.~Andersen}
  \affiliation{Department of Physics, McGill University, 3600 rue University, Montr\'eal, QC H3A 2T8, Canada}
  \affiliation{Trottier Space Institute, McGill University, 3550 rue University, Montr\'eal, QC H3A 2A7, Canada}
\author[0000-0002-3980-815X]{Shion Andrew}
  \affiliation{MIT Kavli Institute for Astrophysics and Space Research, Massachusetts Institute of Technology, 77 Massachusetts Ave, Cambridge, MA 02139, USA}
  \affiliation{Department of Physics, Massachusetts Institute of Technology, 77 Massachusetts Ave, Cambridge, MA 02139, USA}
\author[0000-0003-3772-2798]{Kevin Bandura}
  \affiliation{Lane Department of Computer Science and Electrical Engineering, 1220 Evansdale Drive, PO Box 6109, Morgantown, WV 26506, USA}
  \affiliation{Center for Gravitational Waves and Cosmology, West Virginia University, Chestnut Ridge Research Building, Morgantown, WV 26505, USA}
\author[0000-0002-3615-3514]{Mohit Bhardwaj}
  \affiliation{McWilliams Center for Cosmology \& Astrophysics, Department of Physics, Carnegie Mellon University, Pittsburgh, PA 15213, USA}
\author[0000-0002-9218-1624]{Kalyani Bhopi}
  \affiliation{Lane Department of Computer Science and Electrical Engineering, 1220 Evansdale Drive, PO Box 6109, Morgantown, WV 26506, USA}
  \affiliation{Center for Gravitational Waves and Cosmology, West Virginia University, Chestnut Ridge Research Building, Morgantown, WV 26505, USA}
\author[0009-0008-9653-6104]{Vadym Bidula}
  \affiliation{Department of Physics, McGill University, 3600 rue University, Montr\'eal, QC H3A 2T8, Canada}
  \affiliation{Trottier Space Institute, McGill University, 3550 rue University, Montr\'eal, QC H3A 2A7, Canada}
\author[0000-0001-8537-9299]{P.~J.~Boyle}
  \affiliation{Department of Physics, McGill University, 3600 rue University, Montr\'eal, QC H3A 2T8, Canada}
\author[0000-0002-1800-8233]{Charanjot Brar}
  \affiliation{National Research Council of Canada, Herzberg Astronomy and Astrophysics, 5071 West Saanich Road, Victoria, BC V9E2E7, Canada}
\author[0009-0001-7664-5142]{Mark Carlson}
  \affiliation{Department of Physics and Astronomy, University of British Columbia, 6224 Agricultural Road, Vancouver, BC V6T 1Z1 Canada}
\author[0000-0003-2047-5276]{Tomas Cassanelli}
  \affiliation{Department of Electrical Engineering, Universidad de Chile, Av. Tupper 2007, Santiago 8370451, Chile}
\author[0009-0007-0757-9800]{Alyssa Cassity}
  \affiliation{Department of Physics and Astronomy, University of British Columbia, 6224 Agricultural Road, Vancouver, BC V6T 1Z1 Canada}
\author[0000-0002-2878-1502]{Shami Chatterjee}
  \affiliation{Cornell Center for Astrophysics and Planetary Science, Cornell University, Ithaca, NY 14853, USA}
\author[0000-0001-6509-8430]{Jean-François Cliche}
  \affiliation{Department of Physics, McGill University, 3600 rue University, Montr\'eal, QC H3A 2T8, Canada}
  \affiliation{Trottier Space Institute, McGill University, 3550 rue University, Montr\'eal, QC H3A 2A7, Canada}
\author[0000-0002-8376-1563]{Alice P.~Curtin}
  \affiliation{Department of Physics, McGill University, 3600 rue University, Montr\'eal, QC H3A 2T8, Canada}
  \affiliation{Trottier Space Institute, McGill University, 3550 rue University, Montr\'eal, QC H3A 2A7, Canada}
\author[0000-0001-7674-5066]{Rachel Darlinger}
  \affiliation{Department of Physics, McGill University, 3600 rue University, Montr\'eal, QC H3A 2T8, Canada}
  \affiliation{Trottier Space Institute, McGill University, 3550 rue University, Montr\'eal, QC H3A 2A7, Canada}
\author[0000-0003-3197-2294]{David R.~DeBoer}
  \affiliation{Department of Astronomy, University of California, Berkeley, CA 94720, United States}
\author[0000-0001-7166-6422]{Matt Dobbs}
  \affiliation{Department of Physics, McGill University, 3600 rue University, Montr\'eal, QC H3A 2T8, Canada}
  \affiliation{Trottier Space Institute, McGill University, 3550 rue University, Montr\'eal, QC H3A 2A7, Canada}
\author[0000-0003-4098-5222]{Fengqiu Adam Dong}
  \affiliation{National Radio Astronomy Observatory, 520 Edgemont Rd, Charlottesville, VA 22903, USA}
\author[0000-0003-3734-8177]{Gwendolyn Eadie}
  \affiliation{David A. Dunlap Department of Astronomy and Astrophysics, 50 St. George Street, University of Toronto, ON M5S 3H4, Canada}
  \affiliation{Department of Statistical Sciences, University of Toronto, 700 University Ave., Toronto, ON M5G 1Z5, Canada}
  \affiliation{Data Sciences Institute, University of Toronto, 700 University Ave., Toronto, ON M5G 1Z5, Canada}
\author[0000-0001-8384-5049]{Emmanuel Fonseca}
  \affiliation{Department of Physics and Astronomy, West Virginia University, PO Box 6315, Morgantown, WV 26506, USA }
  \affiliation{Center for Gravitational Waves and Cosmology, West Virginia University, Chestnut Ridge Research Building, Morgantown, WV 26505, USA}
\author[0000-0002-3382-9558]{B.~M.~Gaensler}
  \affiliation{Department of Astronomy and Astrophysics, University of California Santa Cruz, 1156 High Street, Santa Cruz, CA 95060, USA}
  \affiliation{Dunlap Institute for Astronomy and Astrophysics, 50 St. George Street, University of Toronto, ON M5S 3H4, Canada}
  \affiliation{David A. Dunlap Department of Astronomy and Astrophysics, 50 St. George Street, University of Toronto, ON M5S 3H4, Canada}
\author[0000-0001-6128-3735]{Nina Gusinskaia}
  \affiliation{David A. Dunlap Department of Astronomy and Astrophysics, 50 St. George Street, University of Toronto, ON M5S 3H4, Canada}
  \affiliation{Dunlap Institute for Astronomy and Astrophysics, 50 St. George Street, University of Toronto, ON M5S 3H4, Canada}
  \affiliation{ASTRON, Netherlands Institute for Radio Astronomy, Oude Hoogeveensedijk 4, 7991 PD Dwingeloo, The Netherlands}
\author[0000-0002-1760-0868]{Mark Halpern}
  \affiliation{Department of Physics and Astronomy, University of British Columbia, 6224 Agricultural Road, Vancouver, BC V6T 1Z1 Canada}
\author[0009-0003-3736-2080]{Ian Hendricksen}
  \affiliation{Department of Physics, McGill University, 3600 rue University, Montr\'eal, QC H3A 2T8, Canada}
  \affiliation{Trottier Space Institute, McGill University, 3550 rue University, Montr\'eal, QC H3A 2A7, Canada}
\author[0000-0003-2317-1446]{Jason Hessels}
  \affiliation{Department of Physics, McGill University, 3600 rue University, Montr\'eal, QC H3A 2T8, Canada}
  \affiliation{Trottier Space Institute, McGill University, 3550 rue University, Montr\'eal, QC H3A 2A7, Canada}
  \affiliation{Anton Pannekoek Institute for Astronomy, University of Amsterdam, Science Park 904, 1098 XH Amsterdam, The Netherlands}
  \affiliation{ASTRON, Netherlands Institute for Radio Astronomy, Oude Hoogeveensedijk 4, 7991 PD Dwingeloo, The Netherlands}
\author[0000-0003-3457-4670]{Ronniy C.~Joseph}
  \affiliation{Department of Physics, McGill University, 3600 rue University, Montr\'eal, QC H3A 2T8, Canada}
  \affiliation{Trottier Space Institute, McGill University, 3550 rue University, Montr\'eal, QC H3A 2A7, Canada}
\author[0000-0003-4810-7803]{Jane Kaczmarek}
  \affiliation{26 Dick Perry Avenue, Kensington WA 6151 Australia }
  \affiliation{CSIRO Space \& Astronomy, PO Box 76, Epping, NSW 1710, Australia}
  \affiliation{Dominion Radio Astrophysical Observatory, Herzberg Research Centre for Astronomy and Astrophysics, National Research Council Canada, PO Box 248, Penticton, BC V2A 6J9, Canada}
\author[0000-0001-9345-0307]{Victoria M.~Kaspi}
  \affiliation{Department of Physics, McGill University, 3600 rue University, Montr\'eal, QC H3A 2T8, Canada}
  \affiliation{Trottier Space Institute, McGill University, 3550 rue University, Montr\'eal, QC H3A 2A7, Canada}
\author[0009-0005-7115-3447]{Kholoud Khairy}
  \affiliation{Lane Department of Computer Science and Electrical Engineering, 1220 Evansdale Drive, PO Box 6109, Morgantown, WV 26506, USA}
  \affiliation{Center for Gravitational Waves and Cosmology, West Virginia University, Chestnut Ridge Research Building, Morgantown, WV 26505, USA}
\author[ 0000-0003-1455-2546]{T.~L.~Landecker}
  \affiliation{Dominion Radio Astrophysical Observatory, Herzberg Research Centre for Astronomy and Astrophysics, National Research Council Canada, PO Box 248, Penticton, BC V2A 6J9, Canada}
\author[0000-0003-2116-3573]{Adam E.~Lanman}
  \affiliation{MIT Kavli Institute for Astrophysics and Space Research, Massachusetts Institute of Technology, 77 Massachusetts Ave, Cambridge, MA 02139, USA}
  \affiliation{Department of Physics, Massachusetts Institute of Technology, 77 Massachusetts Ave, Cambridge, MA 02139, USA}
\author[0000-0002-2457-3298]{Albert Wai Kit Lau}
  \affiliation{Dunlap Institute for Astronomy and Astrophysics, 50 St. George Street, University of Toronto, ON M5S 3H4, Canada}
\author[0000-0002-5857-4264]{Mattias Lazda}
  \affiliation{Dunlap Institute for Astronomy and Astrophysics, 50 St. George Street, University of Toronto, ON M5S 3H4, Canada}
  \affiliation{David A. Dunlap Department of Astronomy and Astrophysics, 50 St. George Street, University of Toronto, ON M5S 3H4, Canada}
\author[0000-0002-4209-7408]{Calvin Leung}
  \affiliation{Department of Astronomy, University of California, Berkeley, CA 94720, United States}
  \affiliation{Miller Institute for Basic Research, Stanley Hall, Room 206B, Berkeley, CA 94720}
\author[0000-0002-7164-9507]{Robert A.~Main}
  \affiliation{Department of Physics, McGill University, 3600 rue University, Montr\'eal, QC H3A 2T8, Canada}
  \affiliation{Trottier Space Institute, McGill University, 3550 rue University, Montr\'eal, QC H3A 2A7, Canada}
\author[0000-0002-4279-6946]{Kiyoshi W.~Masui}
  \affiliation{MIT Kavli Institute for Astrophysics and Space Research, Massachusetts Institute of Technology, 77 Massachusetts Ave, Cambridge, MA 02139, USA}
  \affiliation{Department of Physics, Massachusetts Institute of Technology, 77 Massachusetts Ave, Cambridge, MA 02139, USA}
\author[0000-0001-7348-6900]{Ryan Mckinven}
  \affiliation{Department of Physics, McGill University, 3600 rue University, Montr\'eal, QC H3A 2T8, Canada}
  \affiliation{Trottier Space Institute, McGill University, 3550 rue University, Montr\'eal, QC H3A 2A7, Canada}
\author[0000-0002-0772-9326]{Juan Mena-Parra}
  \affiliation{Dunlap Institute for Astronomy and Astrophysics, 50 St. George Street, University of Toronto, ON M5S 3H4, Canada}
  \affiliation{David A. Dunlap Department of Astronomy and Astrophysics, 50 St. George Street, University of Toronto, ON M5S 3H4, Canada}
\author[0000-0001-8845-1225]{Bradley W.~Meyers}
  \affiliation{International Centre for Radio Astronomy Research (ICRAR), Curtin University, Bentley WA 6102 Australia}
  \affiliation{Australian SKA Regional Centre (AusSRC), Curtin University, Bentley WA 6102 Australia}
\author[0000-0002-2551-7554]{Daniele Michilli}
  \affiliation{Laboratoire d'Astrophysique de Marseille, Aix-Marseille Univ., CNRS, CNES, Marseille, France}
\author[0000-0001-8292-0051]{Nikola Milutinovic}
  \affiliation{Department of Physics and Astronomy, University of British Columbia, 6224 Agricultural Road, Vancouver, BC V6T 1Z1 Canada}
\author[0000-0003-0510-0740]{Kenzie Nimmo}
  \affiliation{MIT Kavli Institute for Astrophysics and Space Research, Massachusetts Institute of Technology, 77 Massachusetts Ave, Cambridge, MA 02139, USA}
\author[0000-0002-5254-243X]{Gavin Noble}
  \affiliation{David A. Dunlap Department of Astronomy and Astrophysics, 50 St. George Street, University of Toronto, ON M5S 3H4, Canada}
  \affiliation{Dunlap Institute for Astronomy and Astrophysics, 50 St. George Street, University of Toronto, ON M5S 3H4, Canada}
\author[0000-0002-8897-1973]{Ayush Pandhi}
  \affiliation{David A. Dunlap Department of Astronomy and Astrophysics, 50 St. George Street, University of Toronto, ON M5S 3H4, Canada}
  \affiliation{Dunlap Institute for Astronomy and Astrophysics, 50 St. George Street, University of Toronto, ON M5S 3H4, Canada}
\author[0000-0002-8912-0732]{Aaron B.~Pearlman}
  \affiliation{Department of Physics, McGill University, 3600 rue University, Montr\'eal, QC H3A 2T8, Canada}
  \affiliation{Trottier Space Institute, McGill University, 3550 rue University, Montr\'eal, QC H3A 2A7, Canada}
  \affiliation{Banting Fellow}
  \affiliation{McGill Space Institute Fellow}
  \affiliation{FRQNT Postdoctoral Fellow}
\author[0000-0003-1340-818X]{Jeffrey B.~Peterson}
  \affiliation{McWilliams Center for Cosmology \& Astrophysics, Department of Physics, Carnegie Mellon University, Pittsburgh, PA 15213, USA}
\author[0000-0002-9822-8008]{Emily Petroff}
  \affiliation{Department of Physics, McGill University, 3600 rue University, Montr\'eal, QC H3A 2T8, Canada}
  \affiliation{Trottier Space Institute, McGill University, 3550 rue University, Montr\'eal, QC H3A 2A7, Canada}
  \affiliation{Perimeter Institute of Theoretical Physics, 31 Caroline Street North, Waterloo, ON N2L 2Y5, Canada}
\author[0000-0002-4795-697X]{Ziggy Pleunis}
  \affiliation{Anton Pannekoek Institute for Astronomy, University of Amsterdam, Science Park 904, 1098 XH Amsterdam, The Netherlands}
  \affiliation{ASTRON, Netherlands Institute for Radio Astronomy, Oude Hoogeveensedijk 4, 7991 PD Dwingeloo, The Netherlands}
\author[0000-0002-3430-7671]{Alexander W.~Pollak}
  \affiliation{SETI Institute, 339 Bernardo Ave, Suite 200 Mountain View, CA 94043, USA}
\author[0000-0001-7694-6650]{Masoud Rafiei-Ravandi}
  \affiliation{Department of Physics, McGill University, 3600 rue University, Montr\'eal, QC H3A 2T8, Canada}
\author[0000-0003-3463-7918]{Andre Renard}
  \affiliation{Dunlap Institute for Astronomy and Astrophysics, 50 St. George Street, University of Toronto, ON M5S 3H4, Canada}
\author[0000-0002-4623-5329]{Mawson W.~Sammons}
  \affiliation{Department of Physics, McGill University, 3600 rue University, Montr\'eal, QC H3A 2T8, Canada}
  \affiliation{Trottier Space Institute, McGill University, 3550 rue University, Montr\'eal, QC H3A 2A7, Canada}
\author[0000-0003-3154-3676]{Ketan R.~Sand}
  \affiliation{Department of Physics, McGill University, 3600 rue University, Montr\'eal, QC H3A 2T8, Canada}
  \affiliation{Trottier Space Institute, McGill University, 3550 rue University, Montr\'eal, QC H3A 2A7, Canada}
\author[0000-0001-5504-229X]{Pranav Sanghavi}
  \affiliation{Department of Physics, Yale University, New Haven, CT 06520, USA}
\author[0000-0002-7374-7119]{Paul Scholz}
  \affiliation{Department of Physics and Astronomy, York University, 4700 Keele Street, Toronto, ON MJ3 1P3, Canada}
  \affiliation{Dunlap Institute for Astronomy and Astrophysics, 50 St. George Street, University of Toronto, ON M5S 3H4, Canada}
\author[0000-0002-4823-1946]{Vishwangi Shah}
  \affiliation{Department of Physics, McGill University, 3600 rue University, Montr\'eal, QC H3A 2T8, Canada}
  \affiliation{Trottier Space Institute, McGill University, 3550 rue University, Montr\'eal, QC H3A 2A7, Canada}
\author[0000-0002-6823-2073]{Kaitlyn Shin}
  \affiliation{MIT Kavli Institute for Astrophysics and Space Research, Massachusetts Institute of Technology, 77 Massachusetts Ave, Cambridge, MA 02139, USA}
  \affiliation{Department of Physics, Massachusetts Institute of Technology, 77 Massachusetts Ave, Cambridge, MA 02139, USA}
\author[0000-0003-2631-6217]{Seth R.~Siegel}
  \affiliation{Perimeter Institute of Theoretical Physics, 31 Caroline Street North, Waterloo, ON N2L 2Y5, Canada}
  \affiliation{Department of Physics, McGill University, 3600 rue University, Montr\'eal, QC H3A 2T8, Canada}
  \affiliation{Trottier Space Institute, McGill University, 3550 rue University, Montr\'eal, QC H3A 2A7, Canada}
\author[0000-0003-2828-7720]{Andrew Siemion}
  \affiliation{SETI Institute, 339 Bernardo Ave, Suite 200 Mountain View, CA 94043, USA}
\author[0000-0001-6903-5074]{Jonathan L.~Sievers}
  \affiliation{Department of Physics, McGill University, 3600 rue University, Montr\'eal, QC H3A 2T8, Canada}
  \affiliation{Trottier Space Institute, McGill University, 3550 rue University, Montr\'eal, QC H3A 2A7, Canada}
\author[0000-0002-2088-3125]{Kendrick Smith}
  \affiliation{Perimeter Institute of Theoretical Physics, 31 Caroline Street North, Waterloo, ON N2L 2Y5, Canada}
\author[0009-0003-6054-8035]{David Spear}
  \affiliation{Department of Physics and Astronomy, University of British Columbia, 6224 Agricultural Road, Vancouver, BC V6T 1Z1 Canada}
\author[0000-0001-9784-8670]{Ingrid Stairs}
  \affiliation{Department of Physics and Astronomy, University of British Columbia, 6224 Agricultural Road, Vancouver, BC V6T 1Z1 Canada}
\author[0000-0003-4535-9378]{Keith Vanderlinde}
  \affiliation{David A. Dunlap Department of Astronomy and Astrophysics, 50 St. George Street, University of Toronto, ON M5S 3H4, Canada}
  \affiliation{Dunlap Institute for Astronomy and Astrophysics, 50 St. George Street, University of Toronto, ON M5S 3H4, Canada}
\author[0000-0002-1491-3738]{Haochen Wang}
  \affiliation{MIT Kavli Institute for Astrophysics and Space Research, Massachusetts Institute of Technology, 77 Massachusetts Ave, Cambridge, MA 02139, USA}
  \affiliation{Department of Physics, Massachusetts Institute of Technology, 77 Massachusetts Ave, Cambridge, MA 02139, USA}
\author[0000-0002-4560-5316]{Jacob P.~Willis}
  \affiliation{MIT Kavli Institute for Astrophysics and Space Research, Massachusetts Institute of Technology, 77 Massachusetts Ave, Cambridge, MA 02139, USA}
  \affiliation{Department of Physics, Massachusetts Institute of Technology, 77 Massachusetts Ave, Cambridge, MA 02139, USA}
\author[0000-0002-7076-8643]{Tarik J.~Zegmott}
  \affiliation{Department of Physics, McGill University, 3600 rue University, Montr\'eal, QC H3A 2T8, Canada}
  \affiliation{Trottier Space Institute, McGill University, 3550 rue University, Montr\'eal, QC H3A 2A7, Canada}
\newcommand{\allacks}{
B.C.A. is supported by a Fonds de Recherche du Quebec—Nature et Technologies (FRQNT) Doctoral Research Award
FRB research at WVU is supported by an NSF grant (2006548, 2018490)
M.B is a McWilliams fellow and an International Astronomical Union Gruber fellow. M.B. also receives support from the McWilliams seed grant.
A.P.C is a Vanier Canada Graduate Scholar
M.D. is supported by a CRC Chair, NSERC Discovery Grant, and CIFAR.
Fengqiu Adam Dong is supported by an NRAO Jansky Fellowship
G.M.E. is supported by NSERC Discovery Grant and by a Collaborative Resaerch Team grant from the Canadian Statistical Sciences Institute which is supported by NSERC.
E.F. is supported by an NSF grant (2407399).
J.W.T.H. and the AstroFlash research group acknowledge support from a Canada Excellence Research Chair in Transient Astrophysics (CERC-2022-00009); the European Research Council (ERC) under the European Union’s Horizon 2020 research and innovation programme (`EuroFlash'; Grant agreement No. 101098079); and an NWO-Vici grant (`AstroFlash'; VI.C.192.045).
V.M.K. holds the Lorne Trottier Chair in Astrophysics \& Cosmology, a Distinguished James McGill Professorship, and receives support from an NSERC Discovery grant (RGPIN 228738-13).
A.W.K.L is a Dunlap postdoctoral fellow, The Dunlap Institute is funded through an endowment established by the David Dunlap family and the University of Toronto.
C. L. is supported by a Miller Fellowship in the Departments of Astronomy and Physics at UC Berkeley.
K.W.M. holds the Adam J. Burgasser Chair in Astrophysics and is supported by an NSF grant (2018490).
J.M.P. acknowledges the support of an NSERC Discovery Grant (RGPIN-2023-05373).
D.M. acknowledges support from the French government under the France 2030 investment plan, as part of the Initiative d'Excellence d'Aix-Marseille Universit\'e -- A*MIDEX (AMX-23-CEI-088 AMX-21-IET-016).
K.N. is an MIT Kavli Fellow. 
A.P. is funded by the NSERC Canada Graduate Scholarships -- Doctoral program.
A.B.P. is a Banting Fellow, a McGill Space Institute~(MSI) Fellow, and a Fonds de Recherche du Quebec -- Nature et Technologies~(FRQNT) postdoctoral fellow.
Z.P. is supported by an NWO Veni fellowship (VI.Veni.222.295).
M.W.S. acknowledges support from the Trottier Space Institute Fellowship program.
K.R.S is supported by a Fonds de Recherche du Quebec—Nature et Technologies (FRQNT) Doctoral Research Award
P.S. acknowledges the support of an NSERC Discovery Grant (RGPIN-2024-06266).
V.S. is supported by a Fonds de Recherche du Quebec—Nature et Technologies (FRQNT) Doctoral Research Award
K.S. is supported by the NSF Graduate Research Fellowship Program.
J.L.S. is supported by the Canada-150 programme.
FRB work at UBC is supported by the Canadian Institute for Advanced Researach and an NSERC Discovery Grant. The baseband recorder on the CHIME telescope is funded in part by a Canada Foundation for Innovation John R. Evans Lesaders Fund grant to IHS.
J.W. is supported by the United States Space Force under a MIT Lincoln Laboratory Military Fellowship
}